\documentclass[amssymb,amsmath,aps,showpacs,floatfix,nofootinbib,showpacs,12pt]{revtex4}
\usepackage{epsfig}
\usepackage{color}
\usepackage{graphicx}

\begin{document}
\title{Cosmic $e^\pm$, $\bar p$, $\gamma$ and neutrino rays in \\leptocentric dark matter models
}

\author{$^{1,2}$Xiao-Jun Bi, $^{1,3}$Xiao-Gang He, $^4$Ernest Ma, $^2$Juan Zhang}
\affiliation{
$^1$Center for High Energy Physics, Peking University, Beijing 100871\\
$^2$Laboratory of Particle Astrophysics, Institute of High Energy Physics,Chinese Academy of Sciences, Beijing 100049\\
$^3$Department of Physics, Center for Theoretical
Sciences, and LeCospa Center, National Taiwan University,
Taipei\\
$^4$Department of Physics, University of California, Riverside,
California}

\begin{abstract}

Dark matter annihilation is one of the leading explanations for the
recently observed $e^\pm$ excesses in cosmic rays by PAMELA, ATIC,
FERMI-LAT and HESS. Any dark matter annihilation model proposed to
explain these data must also explain the fact that PAMELA data show
excesses only in $e^\pm$ spectrum but not in anti-proton. It is
interesting to ask whether the annihilation mode into anti-proton is
completely disallowed or only suppressed at low energies. Most
models proposed have negligible anti-protons in all energy ranges.
We show that the leptocentric $U(1)_{B-3L_i}$ dark matter model can explain the
$e^\pm$ excesses with suppressed anti-proton mode at low energies,
but at higher energies there are sizable anti-proton excesses. Near
future data from PAMELA and AMS can provide crucial test for this
type of models. Cosmic $\gamma$ ray data can further rule out some of the
models. We also show that this model has interesting cosmic neutrino
signatures.

\end{abstract}


\maketitle

\section{Introduction}

Recently several experiments have reported $e^\pm$ excesses in
cosmic ray (CR) energy spectrum. Last year the PAMELA collaboration
reported $e^+$ excess in the CR energy spectrum from 10 to
100 GeV, but observed no anti-proton excess~\cite{pamela-e,pamela-p}
compared with predictions from CR physics.
These results are compatible with the previous HEAT and AMS01
experiments (e.g., Ref. \cite{heat,ams}) but with higher precision.
Shortly after the ATIC and PPB-BETS balloon experiments have
reported excesses in the $e^+ + e^-$ spectrum between 300 and 800
GeV~\cite{atic,ppb-bets}. The ATIC data show a sharp falling in the
energy spectrum around 600 GeV. Newly published result from
FERMI-LAT collaboration also shows excesses in the $e^+ + e^-$
energy spectrum above the background~\cite{fermi}. However, the
spectrum is softer than that from ATIC. In addition, the HESS
collaboration has inferred a flat but statistically limited
$e^++e^-$ spectrum between 340~GeV and 1~TeV~\cite{hess1} which
falls steeply above 1~TeV~\cite{hess2}. These observational data
have generated a lot of excitement as the excesses may be explained
by dark matter (DM) annihilation or decay with appropriate properties
\cite{darkmatter,df,lepto,bi-he,review}.

If DM annihilation is responsible for the observed data, a lot of
information can be extracted about DM properties\cite{darkmatter,df,lepto,bi-he,review}. The mass of the
annihilating DM serves as the cut off scale of the $e^\pm$ spectrum,
the lepton spectra must have a cut off energy at the DM mass $m_D$.
The FERMI-LAT and HESS data would require that the DM mass to be
around  a few TeV. The DM belongs to the weakly interacting massive
particle (WIMP) category. The fact that PAMELA did not see
anti-proton excess indicates that the DM is hadrophobic or
leptophilic. At least hadronic annihilation modes are suppressed.
Also since the same DM annihilation rate producing the $e^\pm$
excesses is related to the annihilation rate producing the
cosmological relic DM density, 23\% of the energy budget of our
universe, the annihilation rate is constrained. The latter requires
that the thermally averaged annihilation rate $\langle \sigma v \rangle$ to be
$3\times 10^{-26}$ cm$^3$s$^{-1}$. If this annihilation rate is used
to calculate the $e^\pm$ spectrum, it is too small by a factor of
100 to 1000. There is the need to boost up the spectrum with a boost
factor $B$ of 100 to 1000. Any DM model proposed to explain the data
should be able to produce the large boost factor too.

To suppress the cosmic anti-protons leptophilic DM models are
proposed by several authors \cite{lepto,bi-he}. In this type of models the DM only interacts
with leptons in the standard model (SM). However, these DM models are
hard to be tested since DM does not interact with hadrons at all.
Another widely considered $U(1)$ extension of the SM
is the gauged B-L. However, if DM interacts with the SM with the gauged
$U(1)_{B-L}$ interaction
too much anti-protons will be produced and therefore in conflict with PAMELA data.
In this work, we propose a particle
physics DM model, the gauged~\cite{b-3t} $U(1)_{B-3L_i}$ DM model, to explain
the $e^\pm$ excess data. Here B is the baryon number and $L_i$ is
one of the $e$, $\mu$ and $\tau$ lepton numbers.

In this model, the DM is a Dirac fermion with nontrivial $U(1)_{B
-3 L_i}$ charge~\cite{b-3t}. The gauge boson of this $U(1)$ group,
$Z^\prime$, mediates DM annihilation into SM particles. This model is not
a pure leptophlic DM model, like the $U(1)_{L_i-L_j}$ model studied in Ref.
\cite{bi-he}. But has larger couplings to leptons than quarks. We refer it as the leptocentric dark model.
The $Z^\prime$ has suppressed couplings to quarks
leading to suppressed anti-proton up to the PAMELA energy reach, but
may lead to excesses at higher energies. With the $Z^\prime$ mass
close to two times of the DM mass, a large boost factor can be
produced through the Breit-Wigner enhancement mechanism~\cite{df,breit-wigner}. The shape
of the $e^\pm$ spectrum helps to further constrain the choice of the lepton
flavor $L_i$ of the $U(1)_{B-3 L_i}$.

All DM annihilation models
proposed to explain $e^\pm$ contribute at certain level to $\gamma$-rays
by final state radiation.
One must check if the model is
compatible with available data. We find that for the $L_i = \tau$ model, the hadronic $\tau$ decays
lead to too much $\gamma$-rays in conflict with data and only the $L_i = L_{e,\mu}$ models can pass the $\gamma$ ray data constraint if one
takes the $e^\pm$ excess spectrum to fit the ATIC or FERMI-LAT data.
It is interesting to note that the $Z^\prime$ has the same couplings for
charged lepton $l_{L_i}$ and neutrino $\nu_{L_i}$. Therefore after fitting
data on $e^\pm$ excesses in CR, the cosmic neutrinos by DM annihilation
from the Galactic center (GC) are predicted. The neutrino signals can be
tested by experiments such as IceCube.

In the next section we will first describe how to treat
the propagation of charged cosmic rays in the Milky Way. Then we
present the leptocentric $U(1)_{B-3L_i}$ DM model in Sec. III. The cosmic $e^\pm$, $\bar{p}$, $\gamma$ and neutrino spectra predicted are given in Sec. IV, V, VI and VII
respectively. Finally we give our conclusions in Sec. VIII.

\section{The propagation models}

Before  detailed studies of the CRs from the leptocentric  $U(1)_{B-3L_i}$ DM model, let us first
briefly describe the CR propagation model
which predicts the positron and anti-proton spectra
to compare with data, and also $\gamma$ and neutrino CRs.

The charged particles propagate
diffusively in the Galaxy due to the scattering with random magnetic
field\cite{Gaisser:1990vg}. The interactions with interstellar media (ISM) and
interstellar radiation field (ISRF) will lead to energy losses of
CRs. For heavy nuclei and unstable nuclei there are fragmentation
processes by collisions with ISM and radioactive decays respectively.
In addition, the overall convection driven by the Galactic wind and
reacceleration due to the interstellar shock will also affect the
propagation processes of CRs. The propagation equation can be
written as\cite{galprop}
\begin{eqnarray} \frac{\partial \psi}{\partial
t} &=&Q({\bf x},p)+\nabla\cdot(D_{xx}\nabla \psi-{\bf
V_c}\psi)+\frac{\partial}{\partial p}p^2D_{pp}\frac{\partial}
{\partial p}\frac{1}{p^2}\psi \nonumber \\
 &-& \frac{\partial}{\partial p}\left[\dot{p}\psi
-\frac{p}{3}(\nabla\cdot{\bf V_c}\psi)\right]-\frac{\psi}{\tau_f}-
\frac{\psi}{\tau_r}, \label{prop}
\end{eqnarray}
where $\psi$ is the density
of cosmic ray particles per unit momentum interval,
$D_{xx}$ is the spatial diffusion coefficient,
${\bf V_c}$ is the convection velocity, $D_{pp}$ is the diffusion
coefficient in momentum space used to describe the reacceleration
process, $\dot{p}\equiv{\rm d}p/{\rm d}t$ is the momentum loss rate,
$\tau_f$ and $\tau_r$ are time scales for fragmentation and
radioactive decay respectively.
Here $Q({\bf x},p)$ is the source term. For the astrophysical sources
they are located within the Galactic disk \cite{galprop,Yin:2008bs}
while for DM annihilation the source is within the whole dark matter halo.
The source term of DM annihilation is given as
\begin{equation}
Q({\bf X},E)=\frac{\langle\sigma v\rangle}{2m_{\chi}^2}
\frac{{\rm d}N}{{\rm d}E} \rho^2({\bf X}),\label{dm_anni_source}
\end{equation}
with $\langle\sigma v\rangle$ the annihilation cross section and $m_{\chi}$
the mass of dark matter particle, ${{\rm d}N}/{{\rm d}E}$
the spectrum per annihilation and
$\rho({\bf X})$ the dark matter density at ${\bf X}$.

The spatial diffusion is regarded as isotropic and described using a
rigidity dependent function
$D_{xx}=\beta D_0\left(\frac{\rho}{\rho_0}\right)^{\delta}$ with
$\rho$ the rigidity of the particles.
The reacceleration is described by the diffusion in momentum space with
the diffusion coefficient $D_{pp}$, which relates with the spatial
diffusion coefficient $D_{xx}$ as shown in \cite{seo94}.
The convection velocity is
cylindrically symmetric and increases linearly with the
height $z$ from the Galactic plane\cite{galprop}.
Finally the energy losses and fragmentations are calculated
according to the interactions between CRs and the interstellar
gas or the interstellar radiation field.

A numerical method to solve Eq. (\ref{prop}) has been developed by Strong
and Moskalenko, known as the GALPROP model
\cite{galprop}. In GALPROP, the realistic
astrophysical inputs such as the interstellar medium (ISM) and
interstellar radiation field (ISRF) are adopted to
calculate the fragmentations and energy losses of CRs. The
parameters are tuned to reproduce all the observational CR spectra at
Earth. The GALPROP model can give very good
descriptions of all kinds of CRs, including the secondaries such as
$e^+$, $\bar{p}$ and diffuse
$\gamma$-rays\cite{galprop,
Moskalenko:2001ya,Strong:1998fr,Strong:2004de}.

In this work we employ GALPROP to calculate the
propagation of CRs. We will adopt
the diffusion $+$ convection (DC) scenario with parameters given
in Ref. \cite{Yin:2008bs}.
The DC model predicts the
 $e^-$, $e^+$, $p$ and $\bar{p}$ spectra with very good agreement with the
data \cite{galprop}.

The main uncertainty of the propagation model comes from the height of the
Galactic diffusion region. If the height is changed the diffusion coefficient
can be adjusted accordingly so that the confinement time of the cosmic particles
within the Galaxy and the secondary products are kept intact.
However, such adjustments
affect the contribution from DM annihilation greatly
\cite{david}. The reason is that the sources of CRs are located within
the thin Galactic disk, which is much smaller than the diffusion region
no matter how to adjust the height.
But the contribution from DM annihilation comes from the whole dark matter halo.
Only when the annihilation takes place within the diffusion region can it
contribute to the observed cosmic ray fluxes. The particles from the region
out of the diffusion region escape freely. Therefore the height affects
the contribution of DM annihilation greatly.

The effects by changing the diffusion height are also different for electrons
and nuclei from DM annihilation.
For electrons and positrons the main effects that affect their spectra
are the energy loss processes, such as the synchrotron radiation or
inverse Compton scattering with the ISRF. Since electrons/positrons
loss energy rapidly, the high energy electrons can not propagate for
long distance, usually smaller than $1$kpc \cite{julien}. On the contrary
the nuclei are mainly affected by the propagation processes
and propagate much longer than the electrons and positrons.
Therefore the diffusion height affects anti-proton contribution by DM
annihilation much more than positrons. In the following we will try
to adjust the diffusion height to suppress the anti-proton contribution
from DM annihilation while the electrons/positrons can fit the PAMELA and
Fermi data at the same time.

In the conventional cosmic ray model the diffusion height $Z_h$ is taken as
4kpc, which can give best fit to all CR data. Considering the data from
diffuse $\gamma$-rays the height should be at least of about $\sim 2$kpc.
In our calculation we will take another set of propagation parameters with the
height $Z_h=2$kpc. All other parameters are adjusted correspondingly to fit
the CR data.

For cosmic $\gamma$- or neutrino-rays, since they can propagate freely in
the Galaxy, one can obtain their fluxes by simply
integrating the source terms along the line of sight (LOS) at any direction.
The flux $\phi(E_{\gamma,\nu})$ for $\gamma$ and neutrino are given by
\begin{equation}
\label{gammanu}
\phi(E_{\gamma,\nu},\psi)=\frac{1}{4\pi}\times\frac{\langle
            \sigma v\rangle}{2m_{\chi}^2}\frac{{\rm d}N}{{\rm d}E_{\gamma,\nu}}
            \times\int_{LOS}\rho^2(l)
            {\rm d}l,
\end{equation}
where $\phi(E_{\gamma, \nu},\psi)$ are the $\gamma$- and neutrino-ray spectra at the
direction $\psi$ and
${{\rm d}N}/{{\rm d}E_{\gamma,\nu}}$ are the $\gamma$- and neutrino-ray spectra
per annihilation, respectively.
For the emission from a diffuse region with
solid angle $\Delta\Omega$, we define the average flux as
\begin{equation}
\label{average}
\phi_{\Delta\Omega}=\frac{1}{\Delta\Omega}\int_{\Delta\Omega}\phi(\psi){\rm
d}\Omega ~.~\,
\end{equation}

In the rest of the sections, we will follow the above prescriptions to calculate the
$e^\pm$, anti-proton, $\gamma$ and neutrino CRs in the $U(1)_{B- 3L_i}$ models.

\section{The leptocentric $U(1)_{B-3L_i}$ dark matter model}

Data from PAMELA show that there are excesses in cosmic $e^\pm$
energy spectrum, but not in anti-proton spectrum. This prompts
several authors to propose leptophilic DM models~\cite{lepto,bi-he}. The simplest
leptophilic model is the gauged $U(1)_{L_i-L_j}$ DM model~\cite{bi-he}. In the
minimal SM, besides the $SU(3)_C\times SU(2)_L\times U(1)_Y$ gauge
groups, the family lepton number difference $L_i - L_j$ can be gauge
without anomaly. The resulting $Z^\prime$ coupling to SM particles
are leptophilic. Among the three possible models, two of the models
$L_e-L_\mu$ and $L_e - L_\tau$ can explain the PAMELA and ATIC
data well, while the $L_\mu - L_\tau$ model can fit PAMELA, FERMI
and HESS data well and disfavor the other two models. This model
leaving very little signal in hadronic mode. The anti-proton excess
is suppressed at a negligible level for all ranges of anti-proton
energies.

With right handed neutrinos, one can gauge the $B-L$ global symmetry
without gauge anomalies. This model has also been studied in the
context of the recent $e^\pm$ excesses. However, in this type of
models, the $Z^\prime$ is non-leptophilic leading to sizable
anti-proton excess in cosmic ray in contradiction with PAMELA data.

To explain the PAMELA data, one does not need to have a pure
leptophilic model. A  modified leptophilic model with sufficiently large lepton fraction
with non-negligible hadron fraction in DM annihilation can be a
viable model. Because the not completely negligible hadronic
fraction, there may be some interesting testable consequences. The
leptocentric $U(1)_{B-3L_i}$ model~\cite{b-3t} is an interesting example of this type.

The $U(1)_{B - 3L_i}$ model can be classified into three different
models with $L_i$ being: a) $L_e$, b) $L_\mu$, and c) $L_\tau$.
Assuming the DM field $\psi$ is a Dirac fermion which couples to
$Z^\prime$ with a charge $a$, the $Z^\prime$ coupling to SM fermions
and DM are given by
\begin{eqnarray}
L_{int} =  Z^\prime_\mu \left [ a g^\prime \bar \psi \gamma^\mu \psi
+ {1\over 3} g^\prime \sum_{j= u,d,c,s,t,b} \bar q_j \gamma^\mu q_j
+ 3 g^\prime (\bar l_i \gamma^\mu l_i + \bar \nu_{L_i} \gamma^\mu_{L_i}
+ \bar \nu_{R_i}\gamma^\mu \nu_{R_i}) \right ]\;,
\end{eqnarray}
where $l_i$ and $\nu_{L_i}$ are one of the lepton generations. $\nu_{R_i}$
is the right handed neutrino which can be light (Dirac type of neutrino) or
heavy (Majorana type of neutrino).
To give a mass to $Z^\prime$, the simplest way is to introduce a SM
singlet Higgs boson $S$ with non-trivial $U(1)_{B-3 L_i}$ charge $b$.
After the singlet develops a non-zero vacuum expectation value $<S> = v_s/\sqrt{2}$,
the $Z^\prime$ mass $m_{Z^\prime}$ is given by $m^2_{Z^\prime} = b^2 g^\prime v_s^2$.

All of the three models a), b) and c) have suppressed hadronic DM
annihilation fractions. This can be easily understood by noticing
that the cross section of DM annihilate to quark final products
$\sigma$ is proportional to $2\times 3\times 3\times (g^\prime/3)^2$
(the first factor 2 takes care of up and down sector, the first 3 is
the number of quark generation, the second 3 is
 the number of color and last factor $g^\prime/3$ is the $Z^\prime$ coupling to
quark), but to charged lepton is proportional to $(3g^\prime)^2$. Therefore the ratio of
final products with quarks to final products with charged leptons is given by
\begin{eqnarray}
R = {\sigma(\sum_j q_j \bar q_j)\over \sigma(l_i \bar l_i)} = {2\over 9}\;.
\end{eqnarray}
Here we have assumed that the phase space with top quarks in the final states are
approximately the same as other lighter quarks and leptons. This approximate is
justified as the DM should have a mass of order TeV to cover the $e^\pm$ excess energy range.

It is clear that this model has suppressed hadronic DM annihilation
fraction, but non-zero at tree level. With the above suppressed
hadronic fraction, all the above three models are allowed by the
PAMELA data. But it is interesting to note, as will be shown later, that because the non-zero
tree level cross section, just beyond the reach of the current
PAMELA data these models predict noticeable anti-proton excesses.
Therefore with higher experimental energy reach, these models will be tested. To
further distinguish different models, one needs to consider more
experimental constraints.

Without carrying out detailed numerical analysis one
can make some judgments on favored models. Considerations
from the $e^\pm$ spectrum shape can help to select preferred models.
To this end we note that model a) would predict a sharp falling in
$e^\pm$ energy around the DM mass because the DM annihilate directly
into a pair of electron and position through the $Z^\prime$ mediation.
This model seems to be favored by the ATIC data, but the more precise
data from FERMI-LAT and HESS do not confirm the sharp falling
feature and therefore disfavored this model. For models b) and c) the
electron and position are secondary final products from $\mu^+\mu^-$
(model b)) and $\tau^+\tau^-$ (model c)) after DM annihilation. The
$e^\pm$ spectrum is much more softer and are favored by FERMI-LAT
and HESS data, but can not fit the ATIC data well.

Cosmic $\gamma$ ray can further distinguish models a), b) and c). Model
c having $\tau^+ \tau^-$ as the primary DM annihilation products
has a large fraction in hadronic final state. There
are a lot of $\pi^0$ which can decay into $\gamma$ pairs which leads
to visible $\gamma$ ray excesses at observable level. The present
data from HESS disfavors this model. On the other hand, for
models a) and b), $\gamma$ ray dominantly comes from final state radiation.
We will show that $\gamma$ ray produced
in this model is just below the current bound.

\section{The $e^\pm$ cosmic ray}

In the $U(1)_{B-3L_i}$ model, the relic density of the dark matter
is controlled by annihilation of $\bar \psi \psi \to Z^{\prime *}
\to \sum_j q_j\bar q_j + l_i \bar l_i + \nu_i \bar \nu_i$, while
the $e^\pm$ excess is mainly determined by $\bar \psi \psi \to
Z^{\prime*} \to l_i \bar l_i$. The interaction rate
$(\sigma
v)_{l_i \bar l_i}$, with the $l_i$ mass neglected is given by
\begin{eqnarray}
(\sigma v)_{l_i \bar l_i} = {9\over \pi} {a^2 g^{\prime
4}m^2_\psi\over (s - m^2_{Z^\prime})^2 + \Gamma^2_{Z^\prime}
m^2_{Z^\prime}}\; ,\label{cs}
\end{eqnarray}
where $v$ is the relative velocity of the two annihilating dark matter and
$s$ is the total DM pair energy squared in the
center of mass frame. $\Gamma_{Z^\prime}$ is the decay width of
the $Z^\prime$ boson. If the $Z^\prime$ mass is below the $\bar
\psi \psi$ threshold which we will assume, the dominant decay
modes of $Z^\prime$ are $Z^\prime \to \sum_j q_j\bar q_j + \bar l_i l_i + \bar \nu_i
\nu_i$, and $\Gamma_{Z^\prime}$ is given by, neglecting final particle
masses
\begin{eqnarray}
\Gamma_{Z^\prime} = {31 g^{\prime 2}\over 24 \pi} m_{Z^\prime}\;.\label{dr}
\end{eqnarray}
The annihilate rate $(\sigma v)_{relic}$ controlling the relic
density is given by $31(\sigma v)_{l_i\bar l_i}/18$.

In the above, we have assumed that there are
only left-handed light neutrinos. If there are light right-handed
neutrinos to pair up with left-handed neutrinos to form Dirac
neutrinos, the factor 31 in these equations should be changed to 40.

Since the relic density of DM is determined by the annihilation
rate $(\sigma v)_{relic}$, the model parameters are thus constrained. The same parameters
will also determine the annihilation rate producing the
$e^\pm$ excesses observed today, which requires a much
larger annihilation rate. A boost factor in the range 100 to 1000 is
necessary. We find that Breit-Wigner resonance enhancement mechanism
works very well in our models if the $Z^\prime$ boson mass is about
two times of the dark matter mass~\cite{df,bi-he,breit-wigner}.

The boost factor in this case comes from the fact that since the
$Z'$ mass $m_{Z'}$ is close to two times of the dark matter mass
$m_\psi$, the annihilation rate is close to the resonant point and is
very sensitive to the thermal kinetic energy of dark matter. To see
this let us rewrite the annihilation rate into a pair of charged
leptons as
\begin{eqnarray}
(\sigma v)_{l_i \bar l_i}  = {9a^2 g^{\prime 4}\over 16 \pi m^2_\psi} {1\over (\delta + v^2/4)^2 + \gamma^2}\;,
\end{eqnarray}
 where we have used the non-relativistic limit of
$s = 4 m^2_\psi + m^2_\psi v^2$, with $\delta$ and $\gamma$
defined as $m^2_{Z^\prime} = 4 m^2_\psi (1-\delta)$, and $\gamma^2
= \Gamma^2_{Z^\prime}(1-\delta)/4 m^2_\psi$.

For thermal dark matter, the velocity $v^2$ is proportional to the
thermal energy of dark matter. It is clear that for small enough
$\delta$ and $\gamma$, the annihilation rate is very sensitive to
the thermal energy and therefore the thermal temperature T. At lower
dark matter thermal energies, the annihilation rate is enhanced
compared with that at higher temperature. This results in a very
different picture of dark matter annihilation than the case for the
usual non-resonant annihilation where the annihilation rate is not
sensitive to dark matter thermal energies. The annihilation process
does not freeze out even after the usual ``freeze out'' time in the
non-resonant annihilation case due to the enhanced annihilation rate
at lower energies. To produce the observed dark matter relic
density, the annihilation rate at zero temperature is required to be
larger than the usual one, and therefor a boost factor. With
appropriate $\delta$ and $\gamma$, a large enough boost factor can
be produced. For our numerical analysis we follow the procedures in
Ref.\cite{bi-he}. 

As already mentioned in the previous section that
if $l_i = e$, the $e^\pm$ is hard, but for $l_i = \mu$ or $\tau$, the $e^\pm$
have to come from $\mu$ and $\tau$ decays leading to softer $e^\pm$ spectra.

\begin{figure}[!ht]
\includegraphics[scale=0.27]{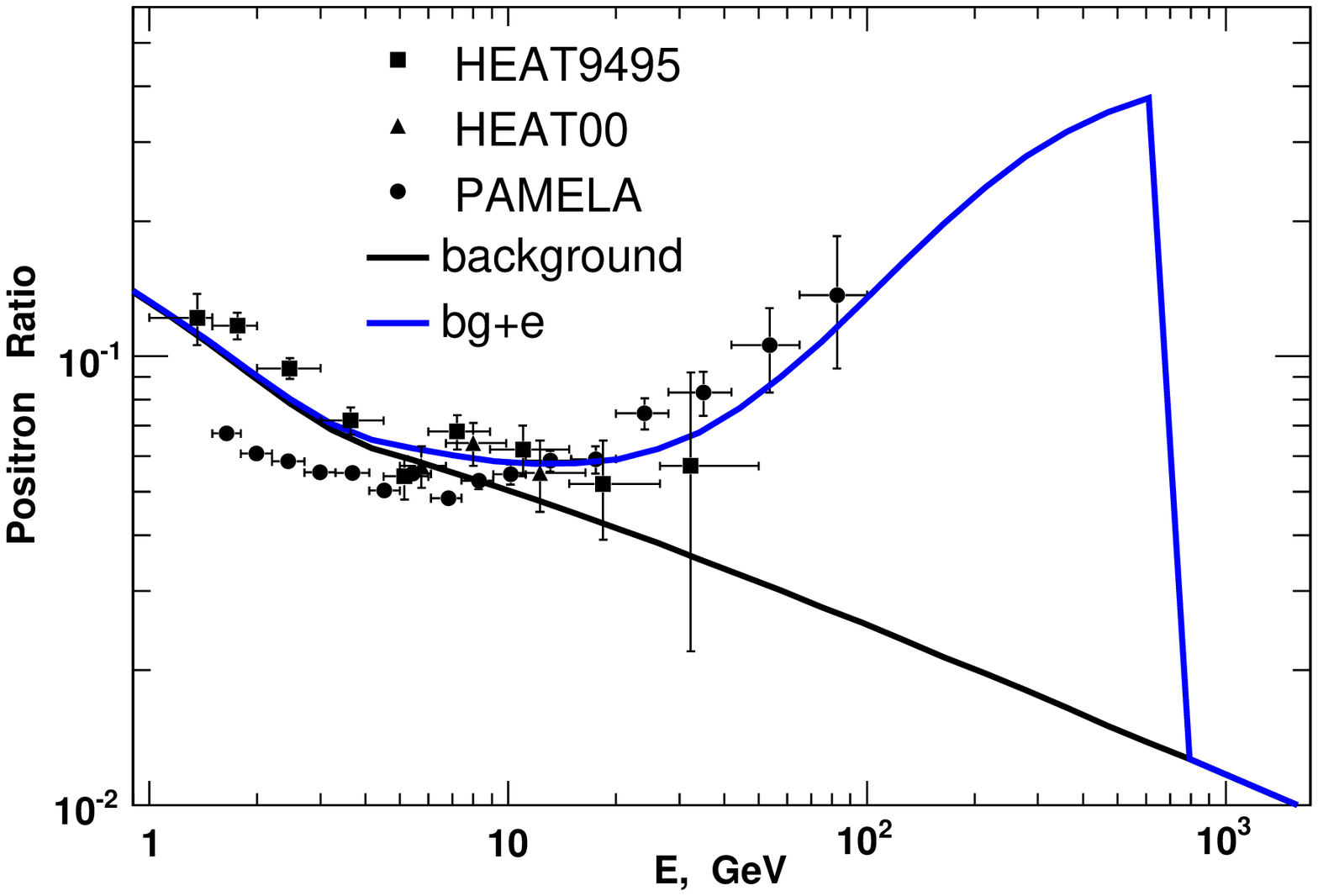}
\includegraphics[scale=0.27]{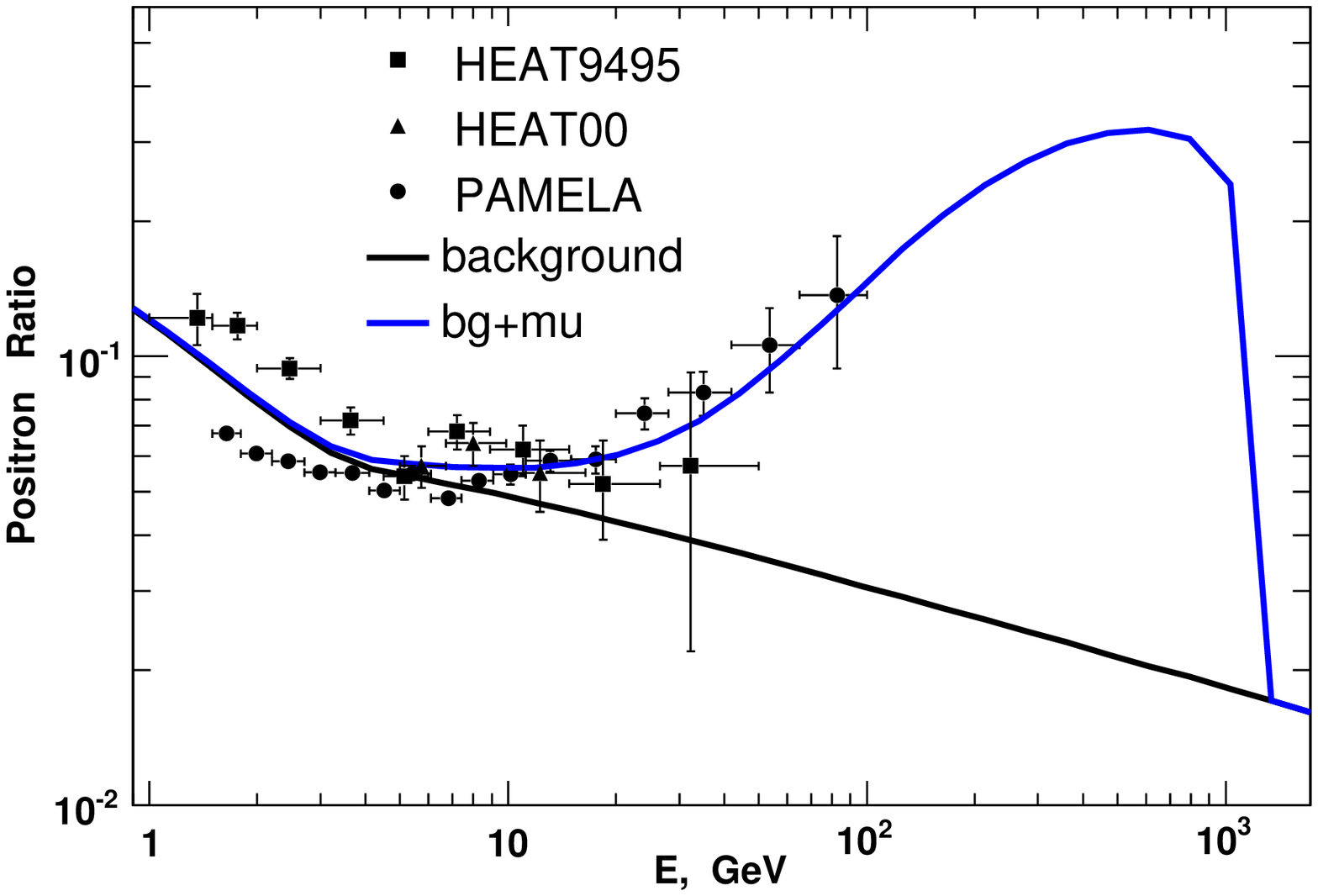}
\includegraphics[scale=0.27]{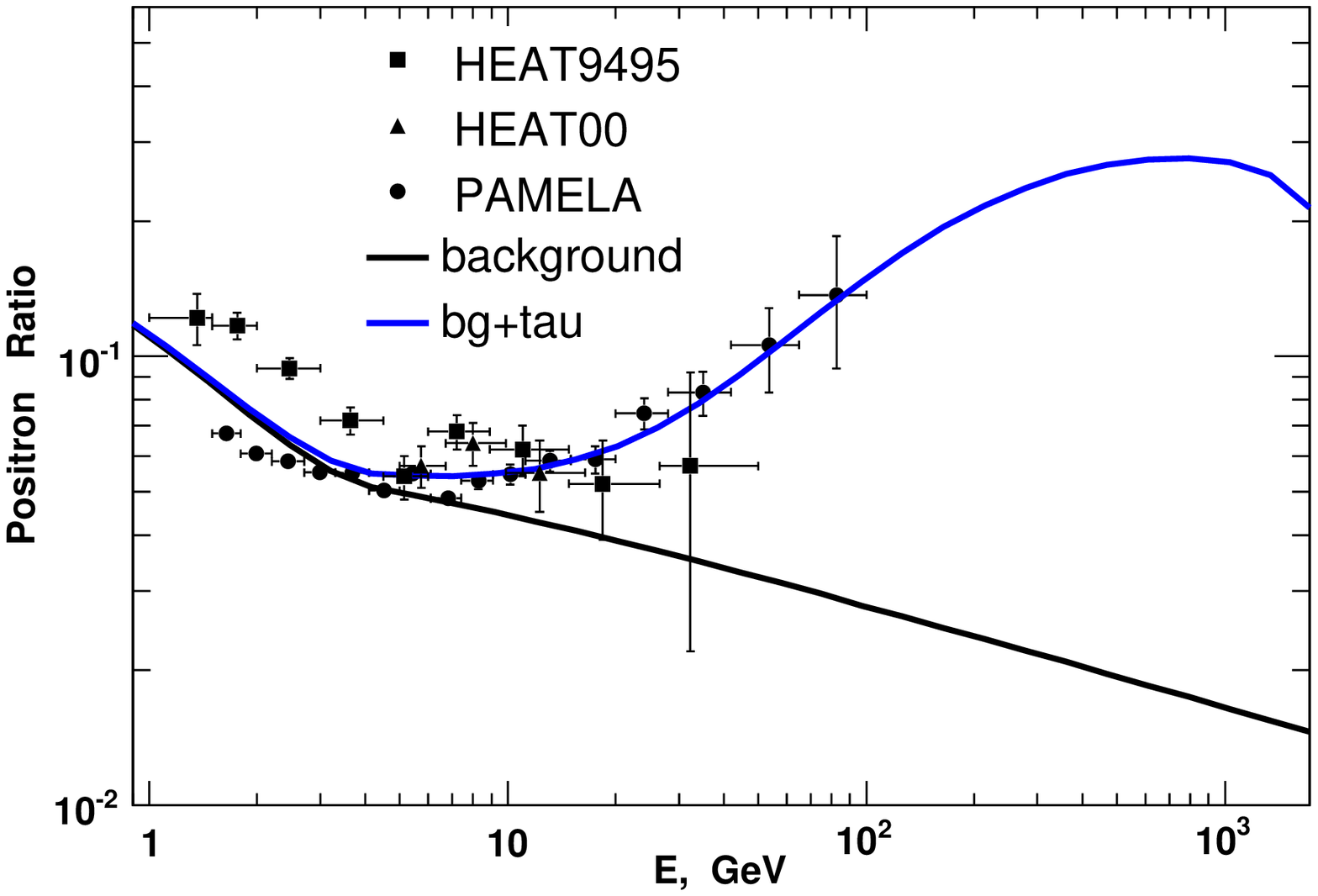}
\includegraphics[scale=0.27]{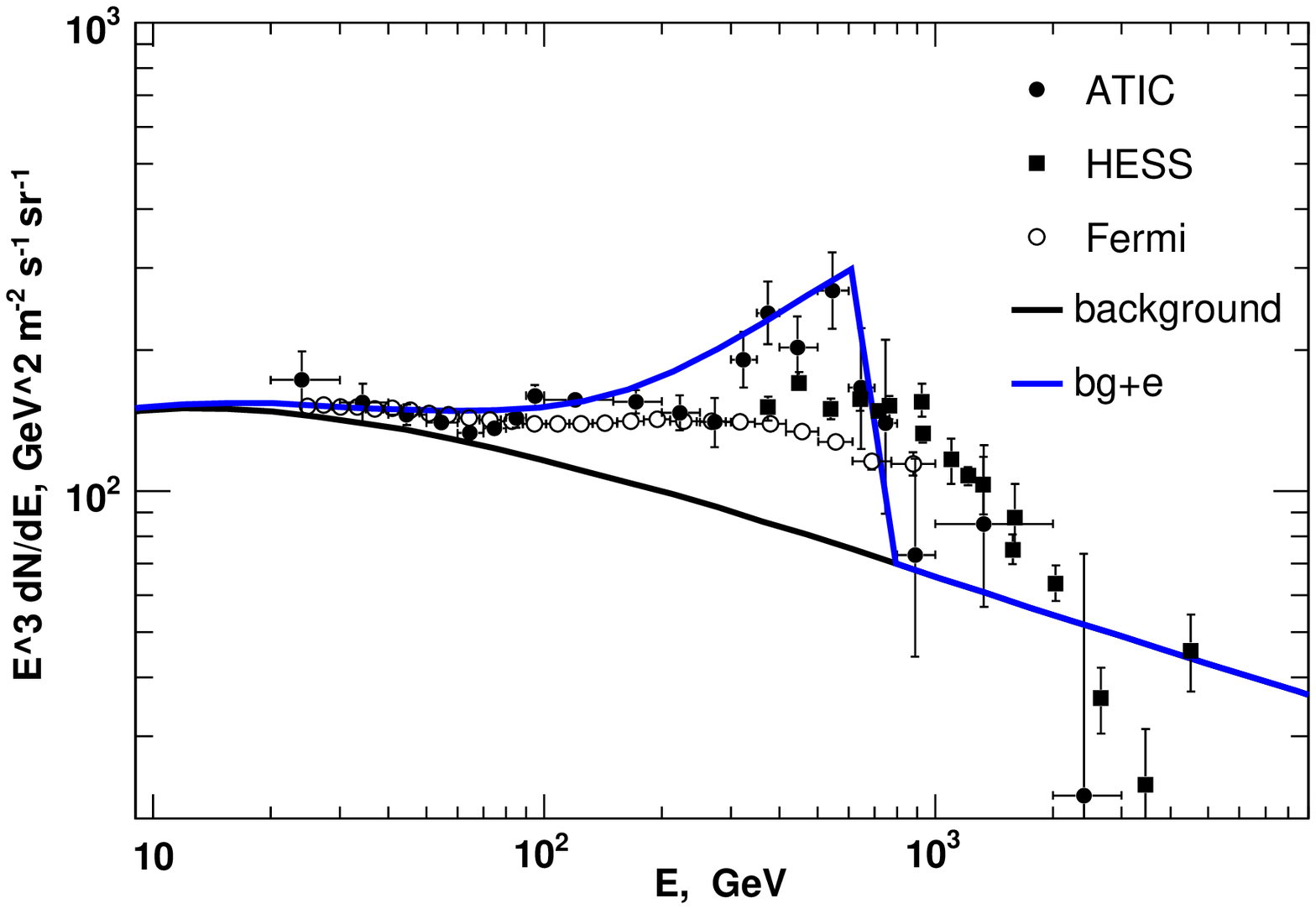}
\includegraphics[scale=0.27]{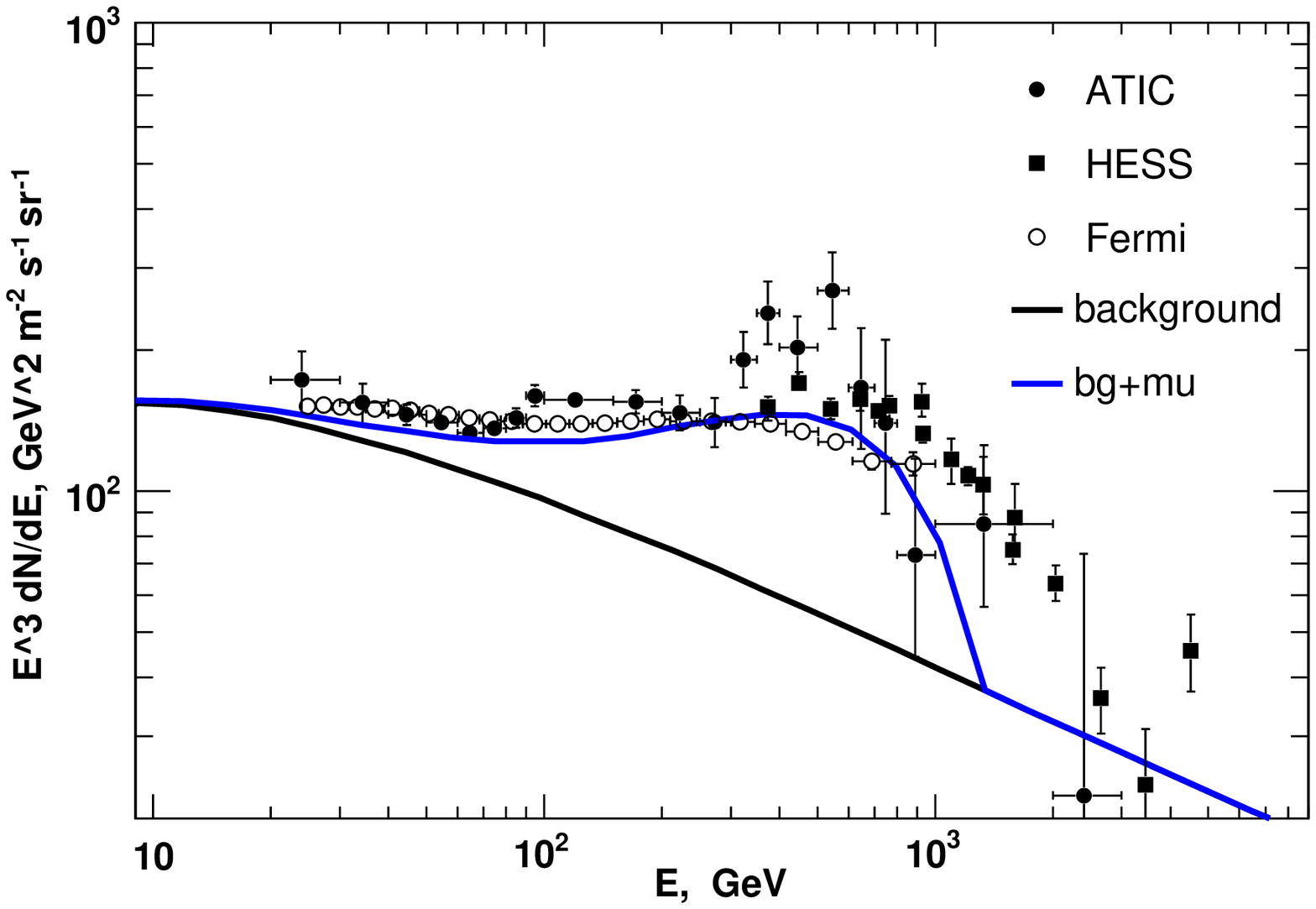}
\includegraphics[scale=0.27]{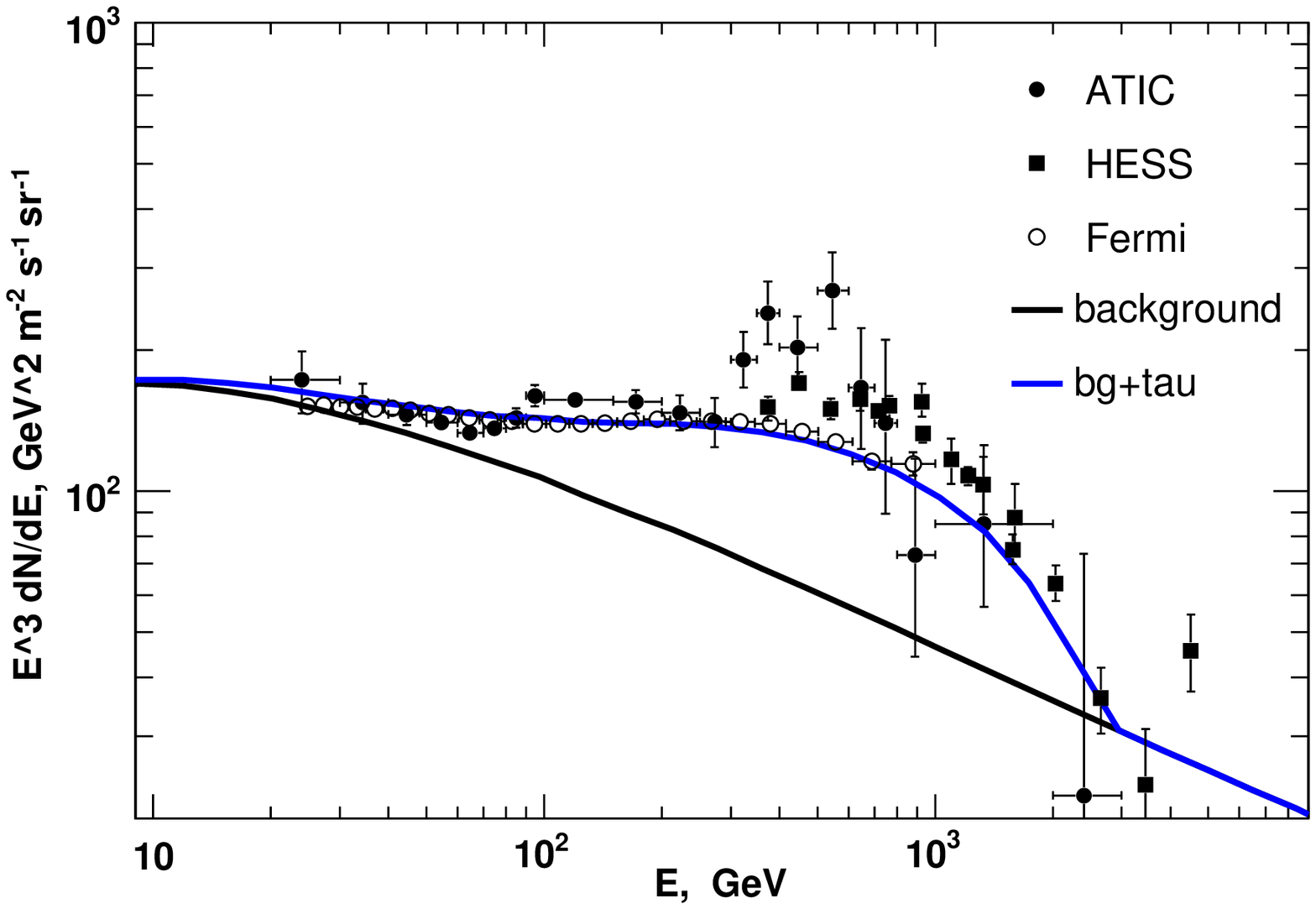}
\caption{The model prediction for $e^\pm$ cosmic ray energy spectrum.
The upper panels show the positron ratio to fit the PAMELA data with
$U(1)_{B-3L_i}$ models. The lower panels show the electron spectrum to
fit the ATIC or Fermi data. The height of the Galactic diffusion region
is taken as 4kpc. }
\label{e4kpc}
\end{figure}

\begin{figure}[!ht]
\includegraphics[scale=0.27]{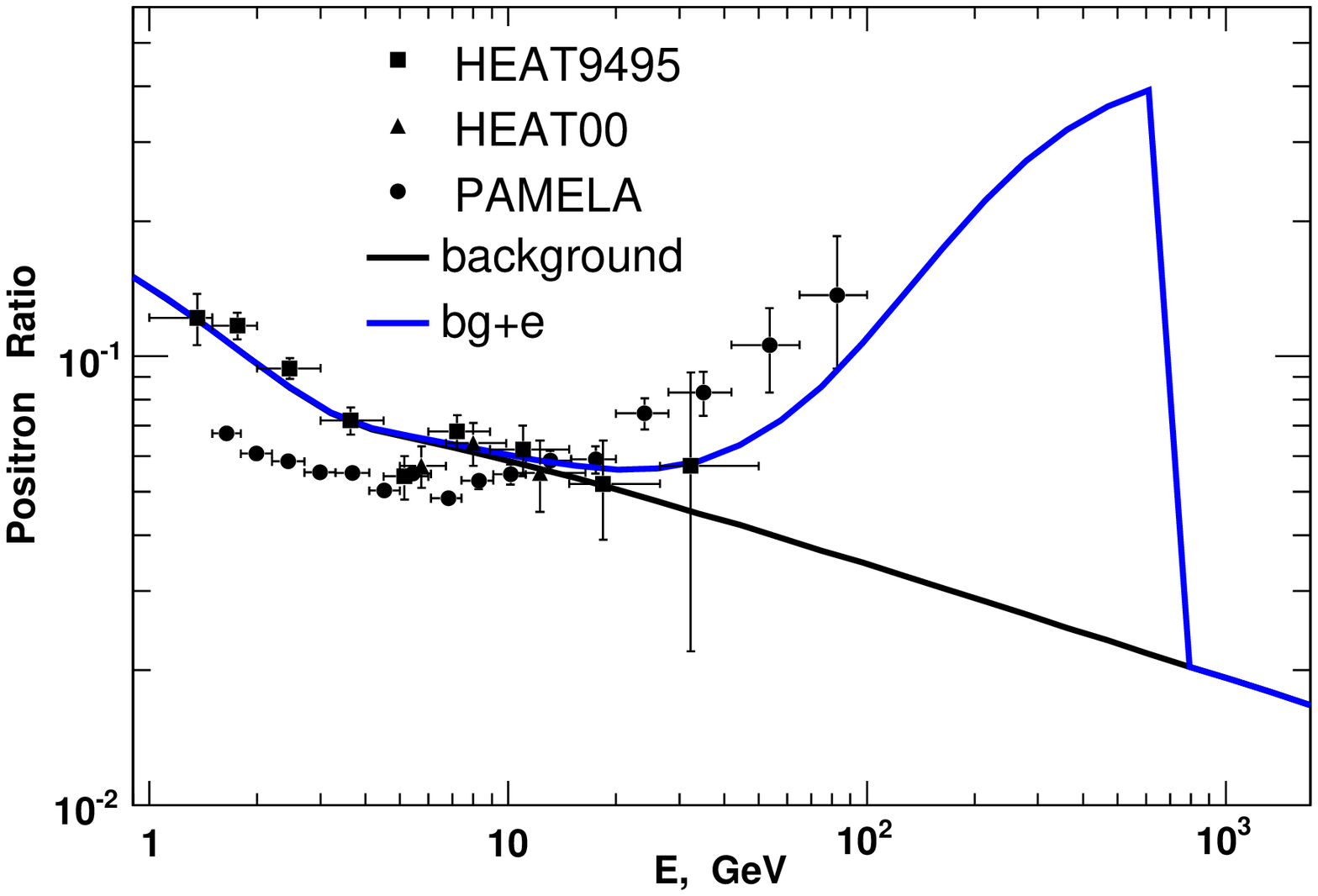}
\includegraphics[scale=0.27]{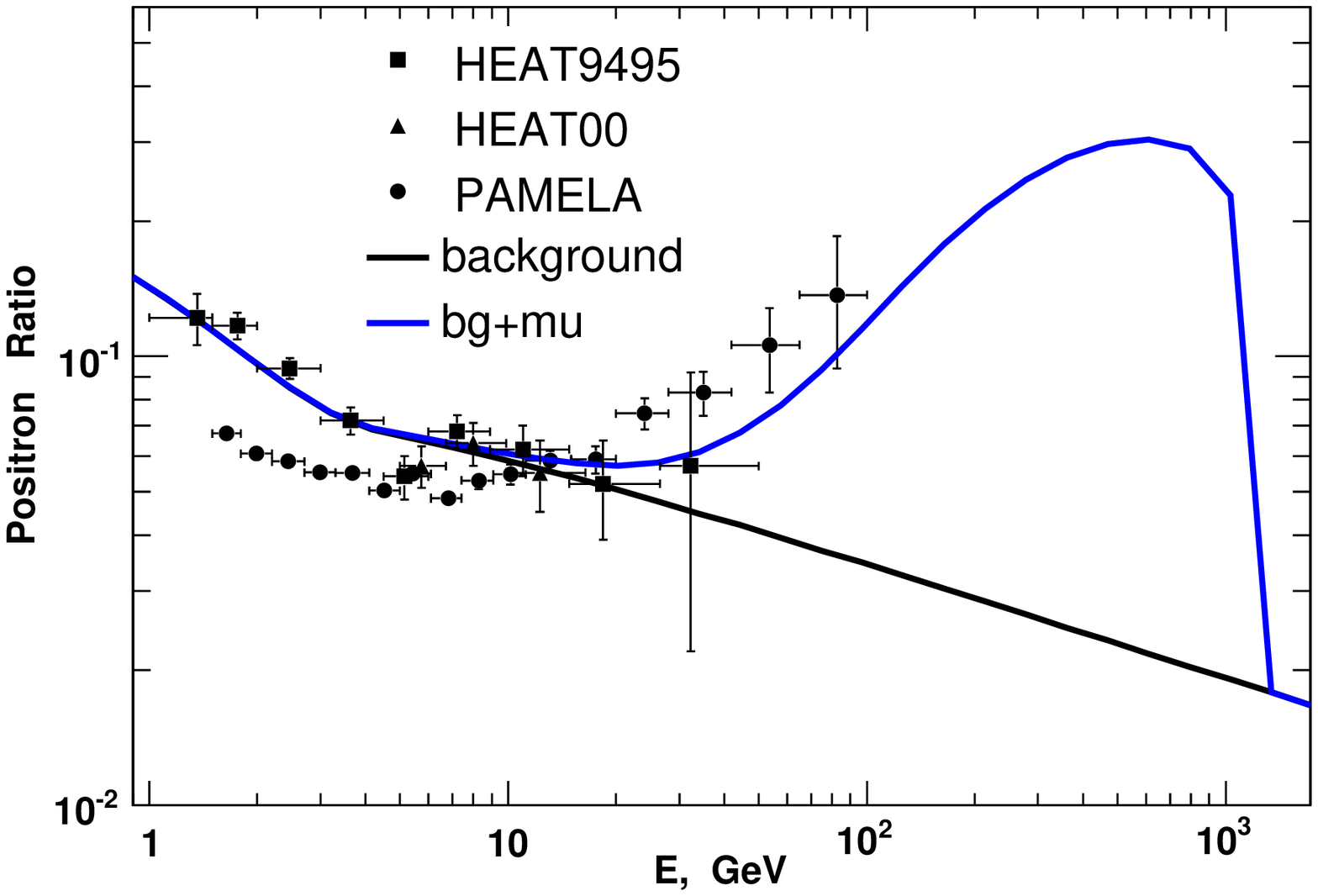}
\includegraphics[scale=0.27]{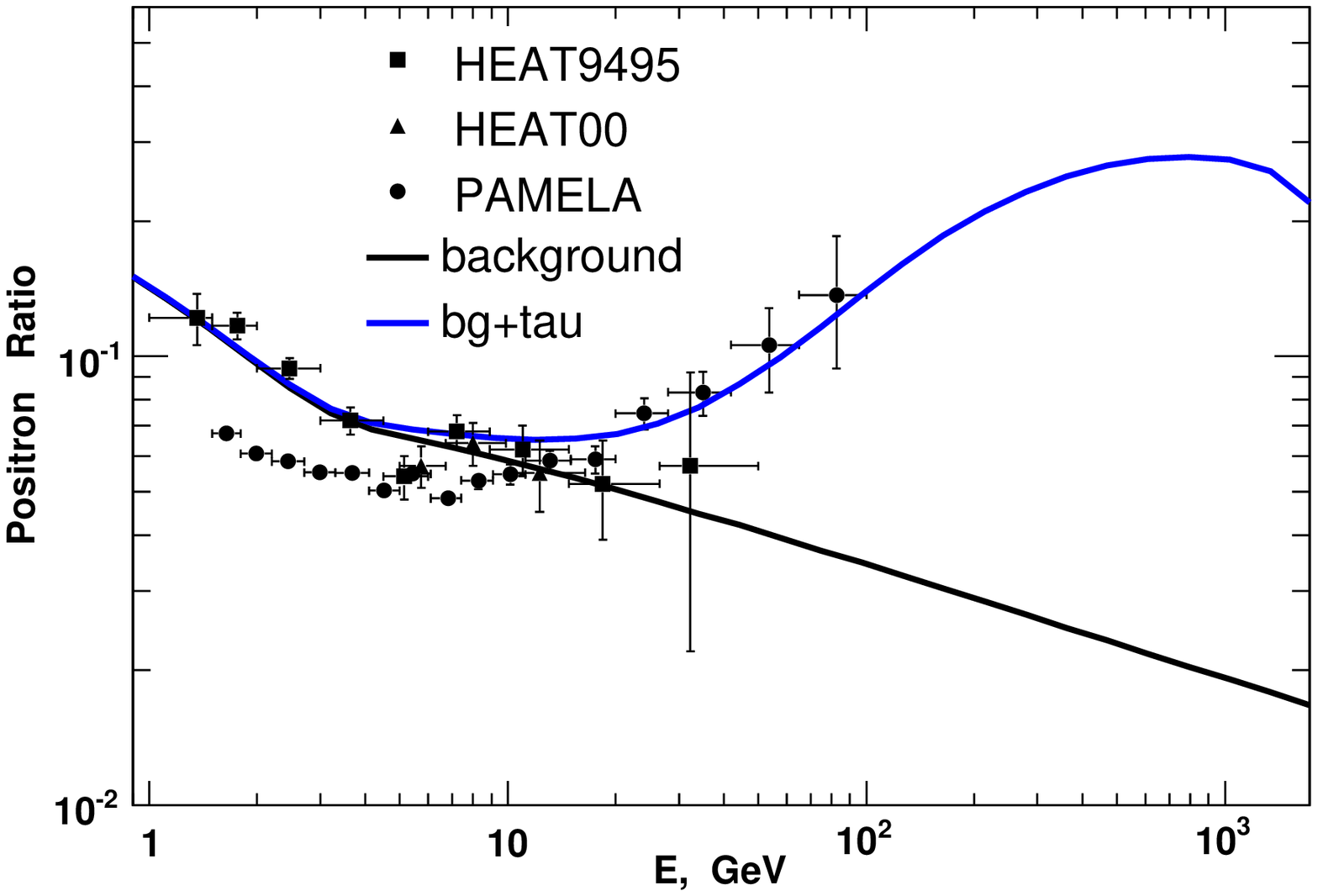}
\includegraphics[scale=0.27]{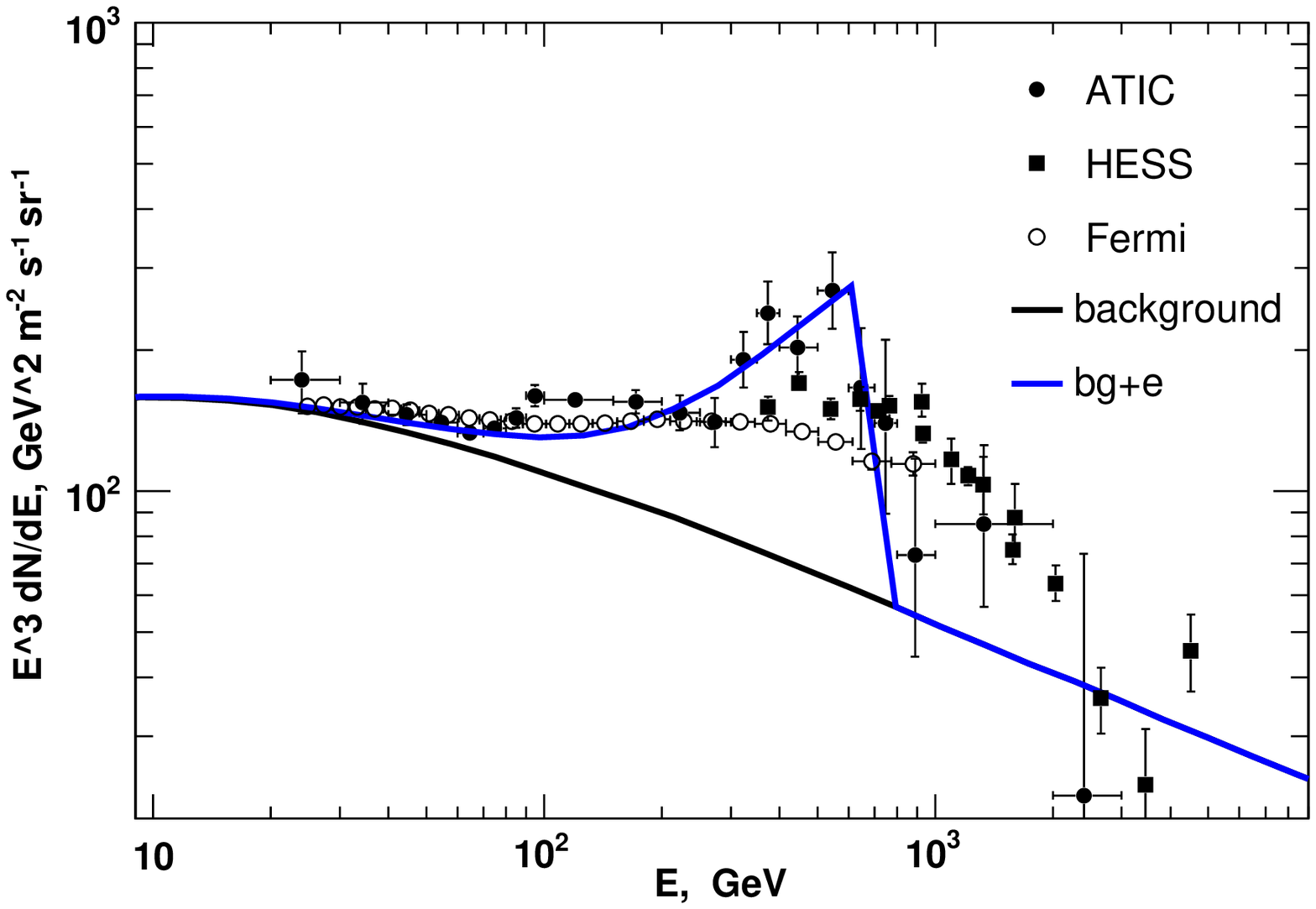}
\includegraphics[scale=0.27]{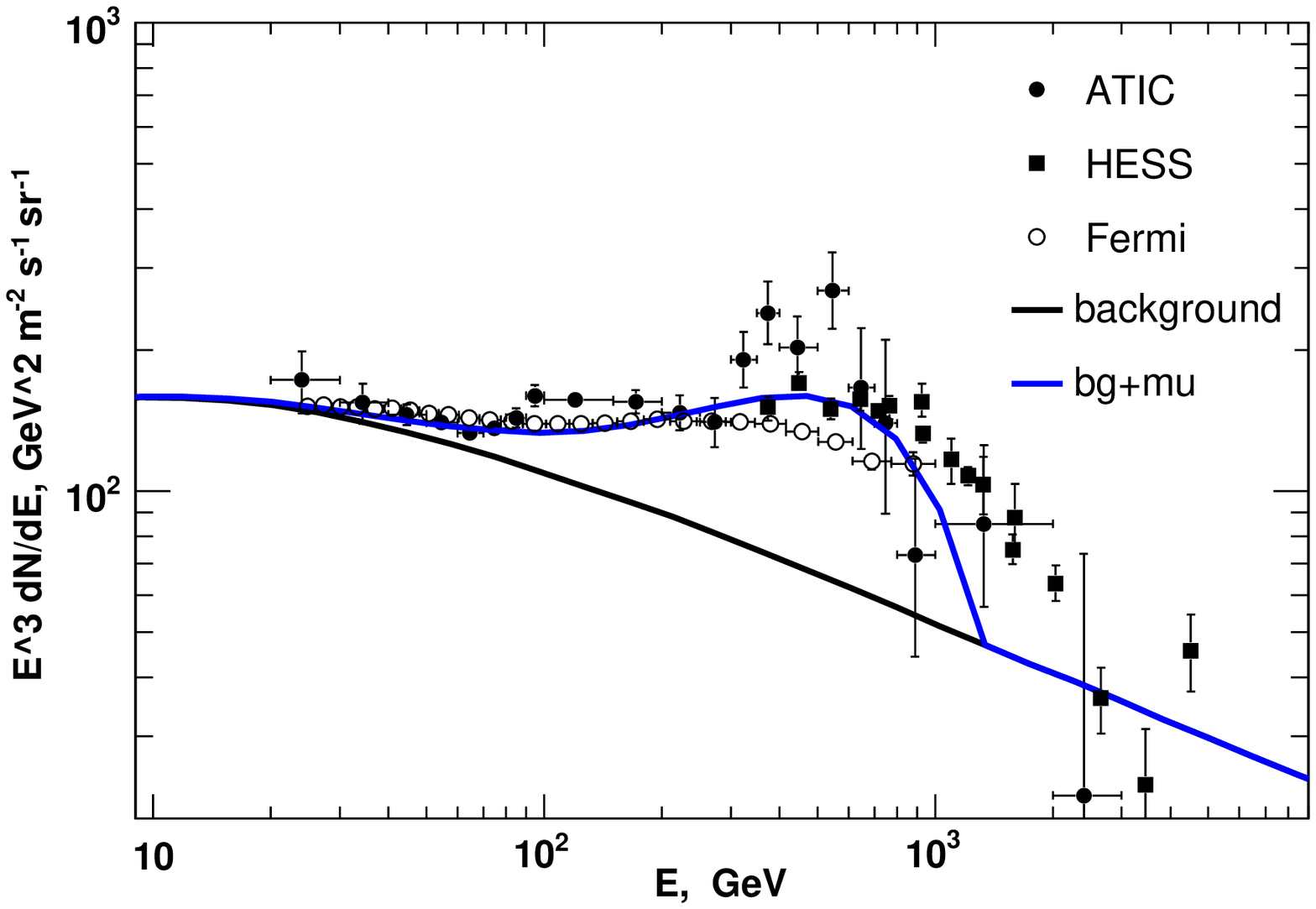}
\includegraphics[scale=0.27]{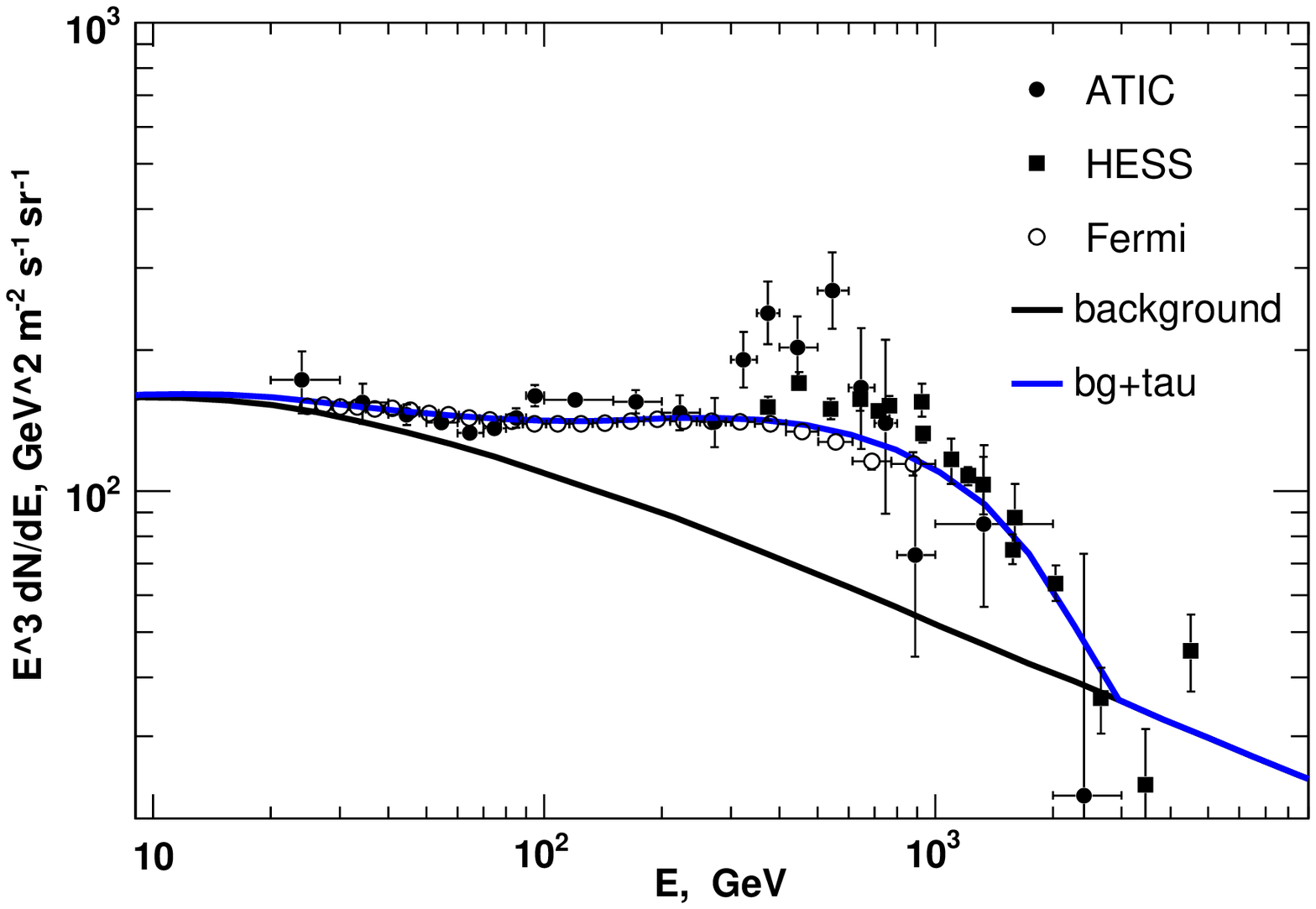}
\caption{The same as Fig. \ref{e4kpc} except the hight of the Galactic
diffusion region taken as 2kpc. }
\label{e2kpc}
\end{figure}

In the Figs. \ref{e4kpc} and \ref{e2kpc} we show our results from fitting
the PAMELA positron ratio and the ATIC or FERMI-LAT electron spectrum
for the $U(1)_{B-3L_e}$, $U(1)_{B-3L_e}$ and $U(1)_{B-3 L_\tau}$ models
taking the diffusion height $Z_h$ 4kpc and 2 kpc respectively.

There are several DM model parameters in the fitting, the DM mass $m_D$, the gauge coupling constant $ag^{\prime 2}$.
Since the Breit-Wigner enhancement for the boost factor forces the $Z^\prime$ mass $m_{Z^\prime}$ to be $2 m_D$, it is not an independent parameter in the models
we are considering. We find that the DM mass for $L_i = e$, $L_i = \mu$ and $L_i = \tau$  are fixed to be in the ranges of several hundred GeV to several TeV in order to cover the
observed ranges of $e^\pm$ excesses. The corresponding values for $a g^{\prime 2}$ are limited to be of order $10^{-5}$.
The masses and the coupling  $ag^{\prime 2}$ used in the figures for three different models are 1 TeV, 1.5 TeV, 3 TeV, and $1.7\times 10^{-5}$, $2.5\times 10^{-5}$ and $5.3 \times 10^{-5}$, respectively.

To see how the astrophysics parameters affect the results, we show results for two different
values for the hight of the diffusion region.
In Fig. \ref{e4kpc} we have taken the hight of the Galactic diffusion
region as 4 kpc, while at Fig. \ref{e2kpc} we take the hight 2 kpc. We have
adjust the DM mass, the annihilation cross section to fit
the data, and also the propagation parameters to
account for all the cosmic ray data. For $l_i = e$, that is for the $U(1)_{B-3L_e}$ model,
the $e^\pm$ spectra from DM annihilation are hard. With appropriate DM mass, from the figures
we can see that this model can fit the PAMELA and ATIC data very well.
For the other two models $l_i = \mu$ (the $U(1)_{B-3 L_\mu}$ model)  or $l_i = \tau$ (the $U(1)_{B - 3L_\tau}$ model),
the $e^\pm$ spectra are from secondary decays of $\mu$ or $\tau$ and therefore are softer. It can be seen from the figures that these two
models can fit the PAMELA and Fermi data.  The $U(1)_{B-3L_\tau}$ model has an even softer $e^\pm$ spectra than those of the $U(1)_{B - 3L_\mu}$
model because $\tau$ has large branching ratios into final states with no $e$ in the final state.

\section{The anti-proton cosmic ray}

The anti-proton from DM annihilation in the $U(1)_{B-3L_i}$ comes from DM
annihilation into quark pairs whose annihilation rate is given by
\begin{eqnarray}
(\sigma v)_{q\bar q} = {1\over 3 \pi} {a^2 g^{\prime 4}m^2_\psi\over (s -
m^2_{Z^\prime})^2 + \Gamma^2_{Z^\prime} m^2_{Z^\prime}}\; .
\end{eqnarray}

\begin{figure}[!ht]
\includegraphics[scale=0.27]{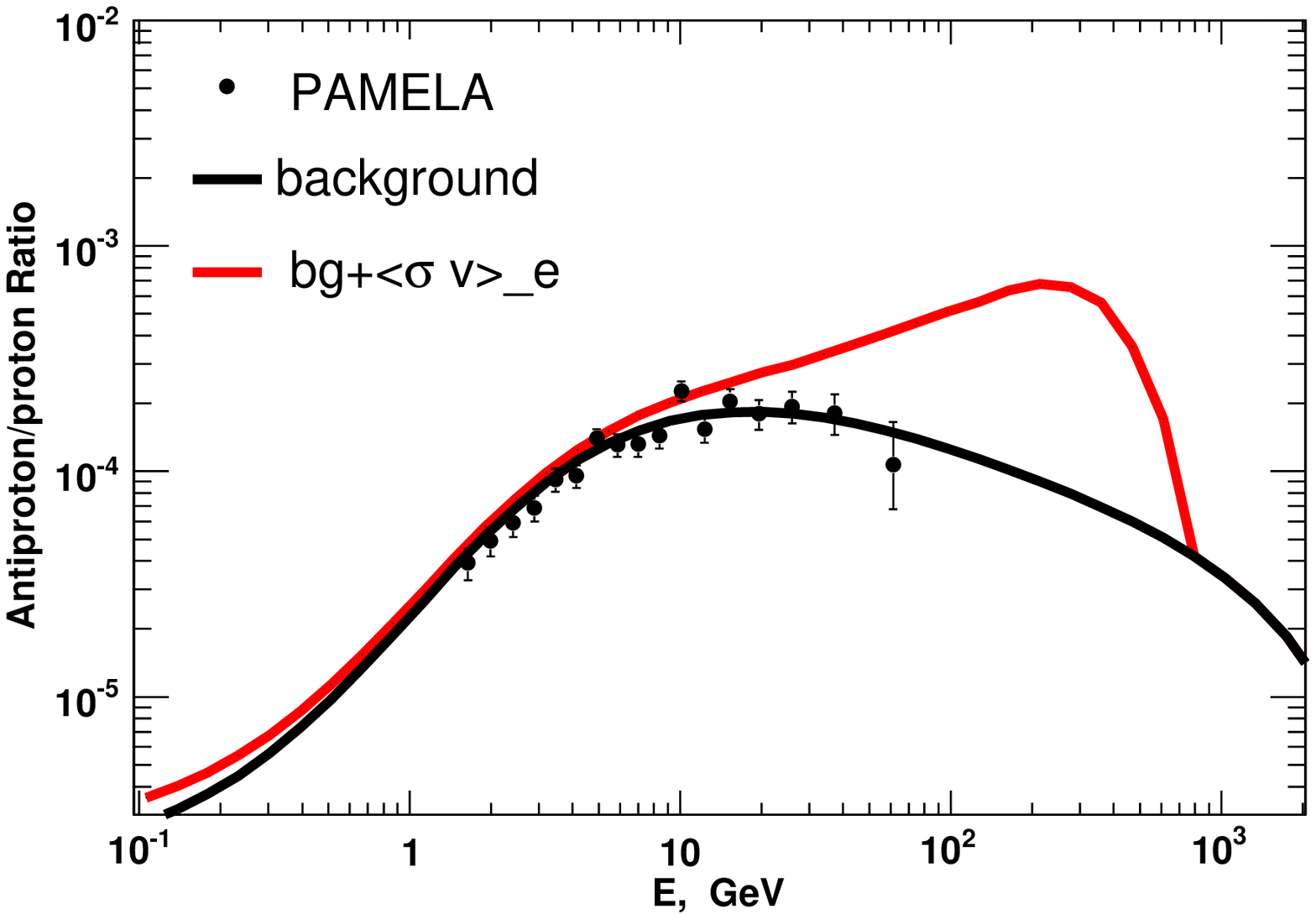}
\includegraphics[scale=0.27]{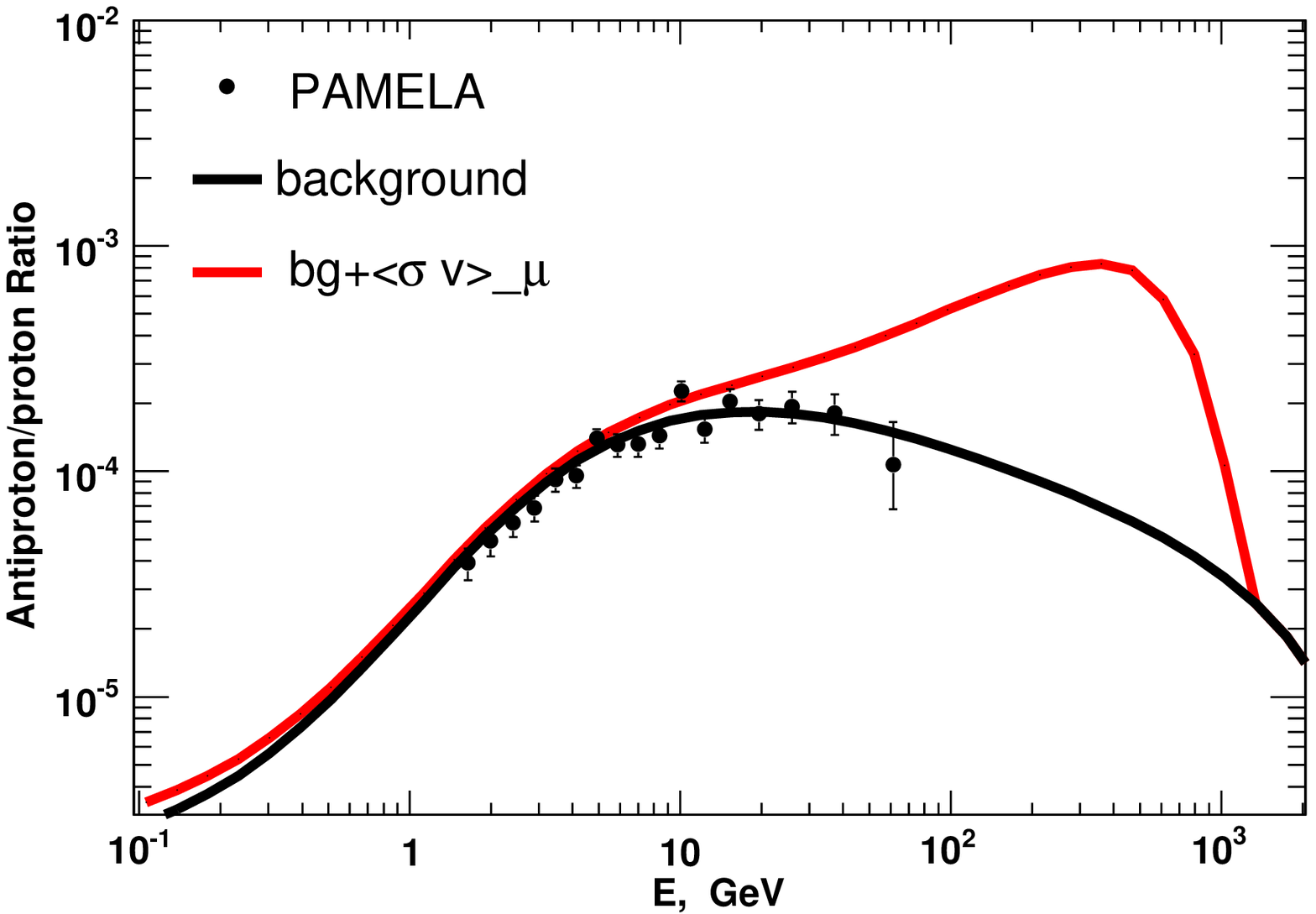}
\includegraphics[scale=0.27]{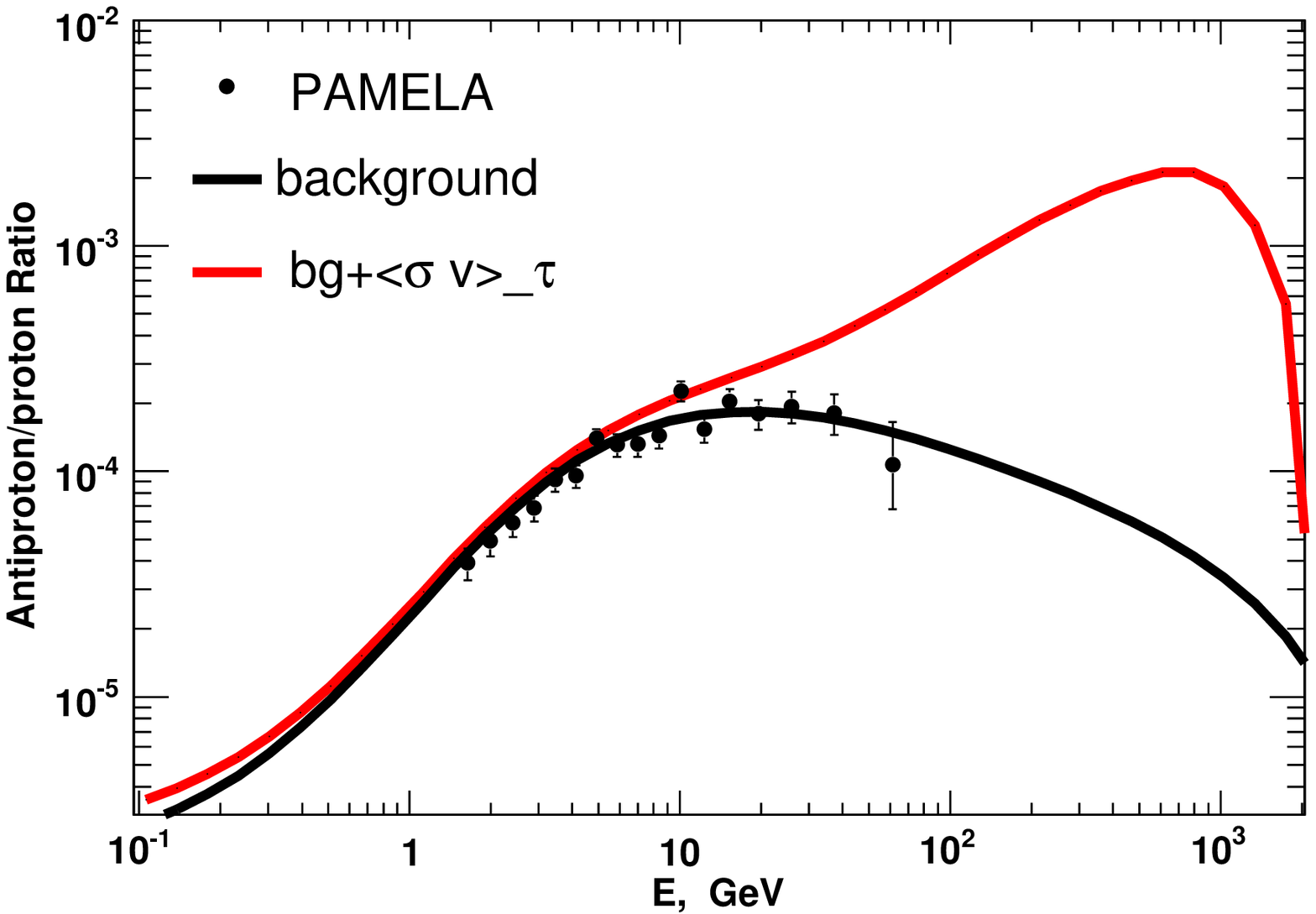}
\caption{The ratio of anti-proton to proton as function of energy predicted
in the B-3e, B-3$\mu$ and B-3$\tau$ models respectively from the left to
the right panels with $Z_h=4$kpc. The cross section of DM annihilation is
normalized to fit the ATIC or Fermi electron spectrum. } \label{anti-proton4kpc}
\end{figure}

\begin{figure}[!ht]
\includegraphics[scale=0.27]{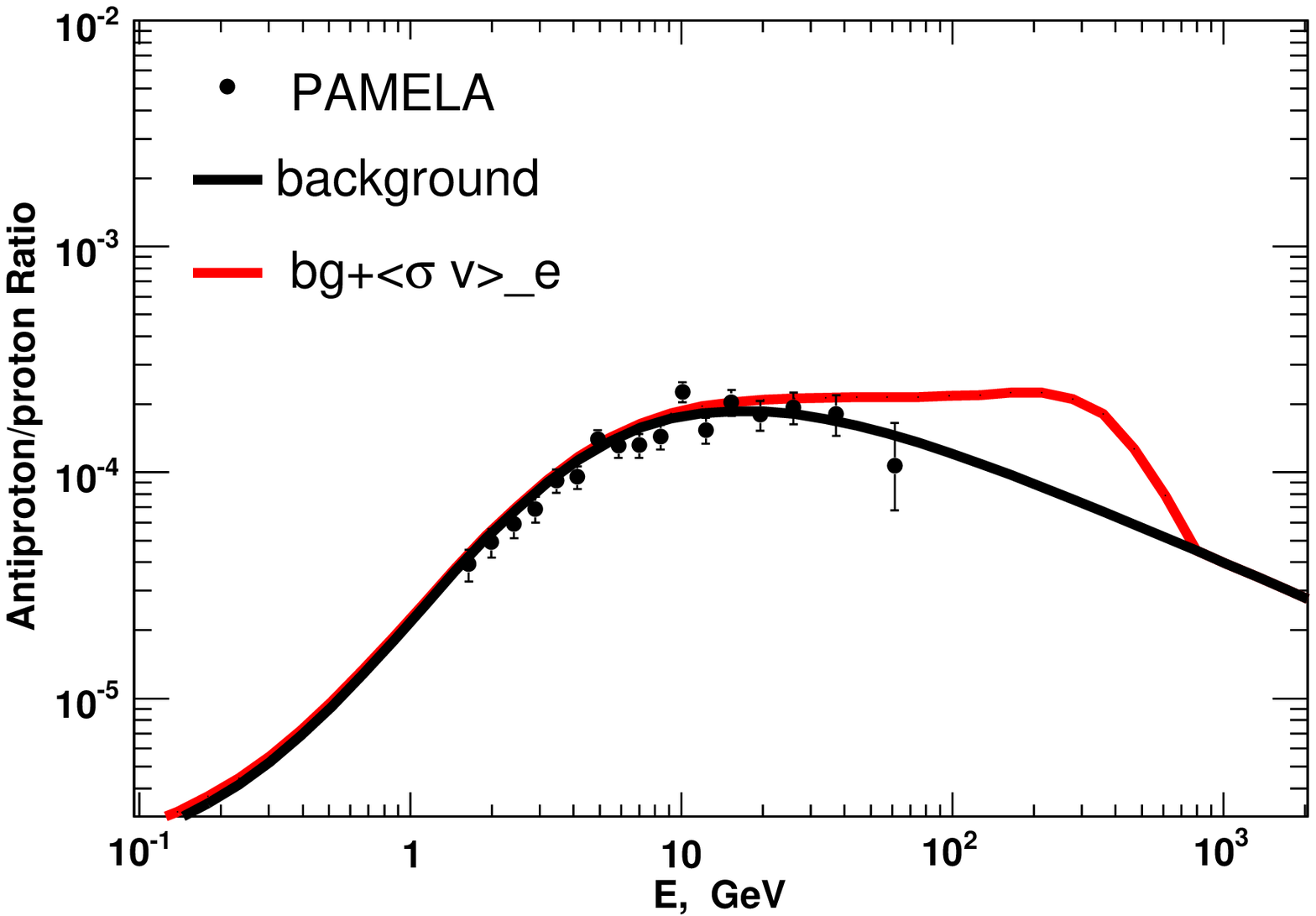}
\includegraphics[scale=0.27]{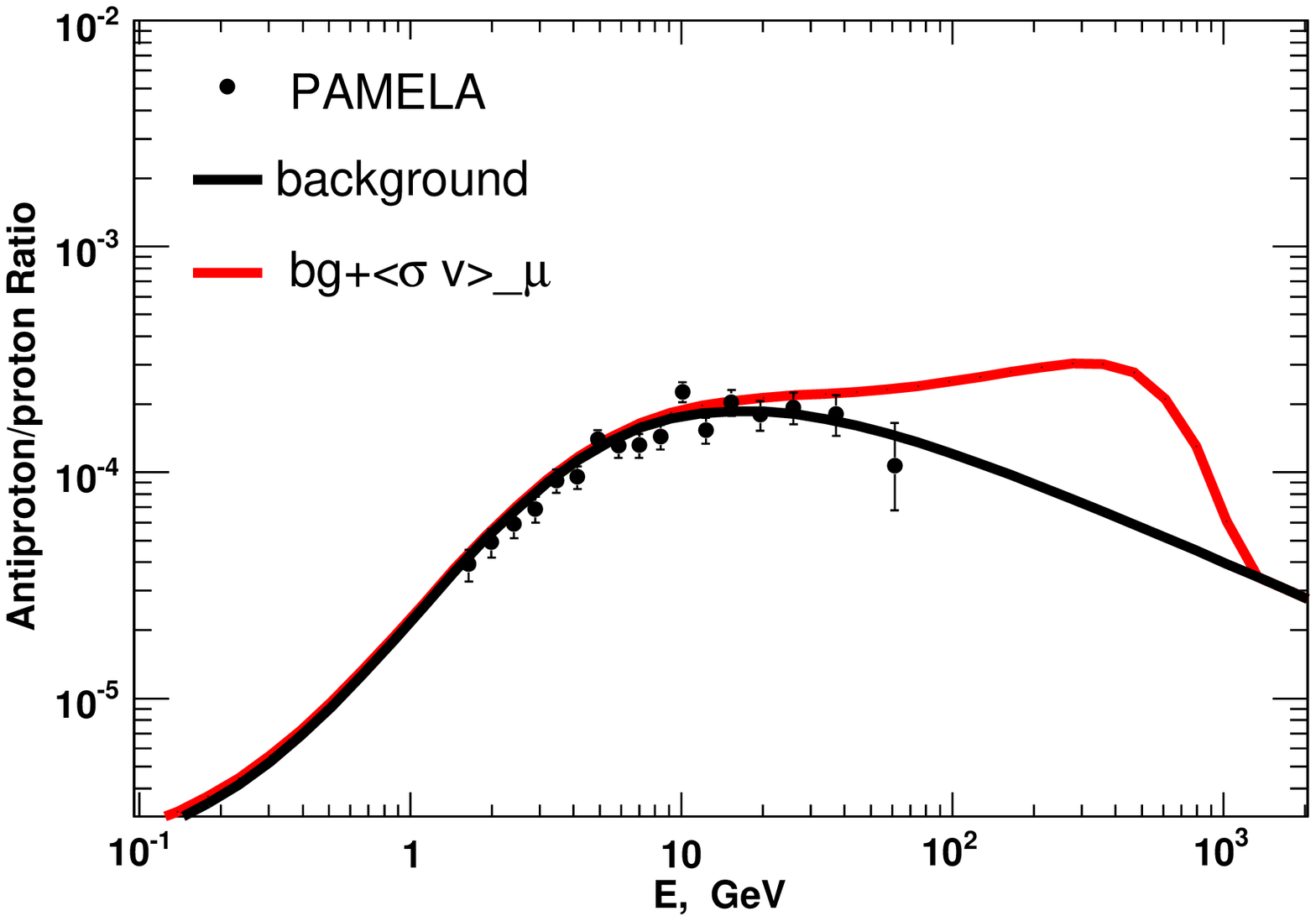}
\includegraphics[scale=0.27]{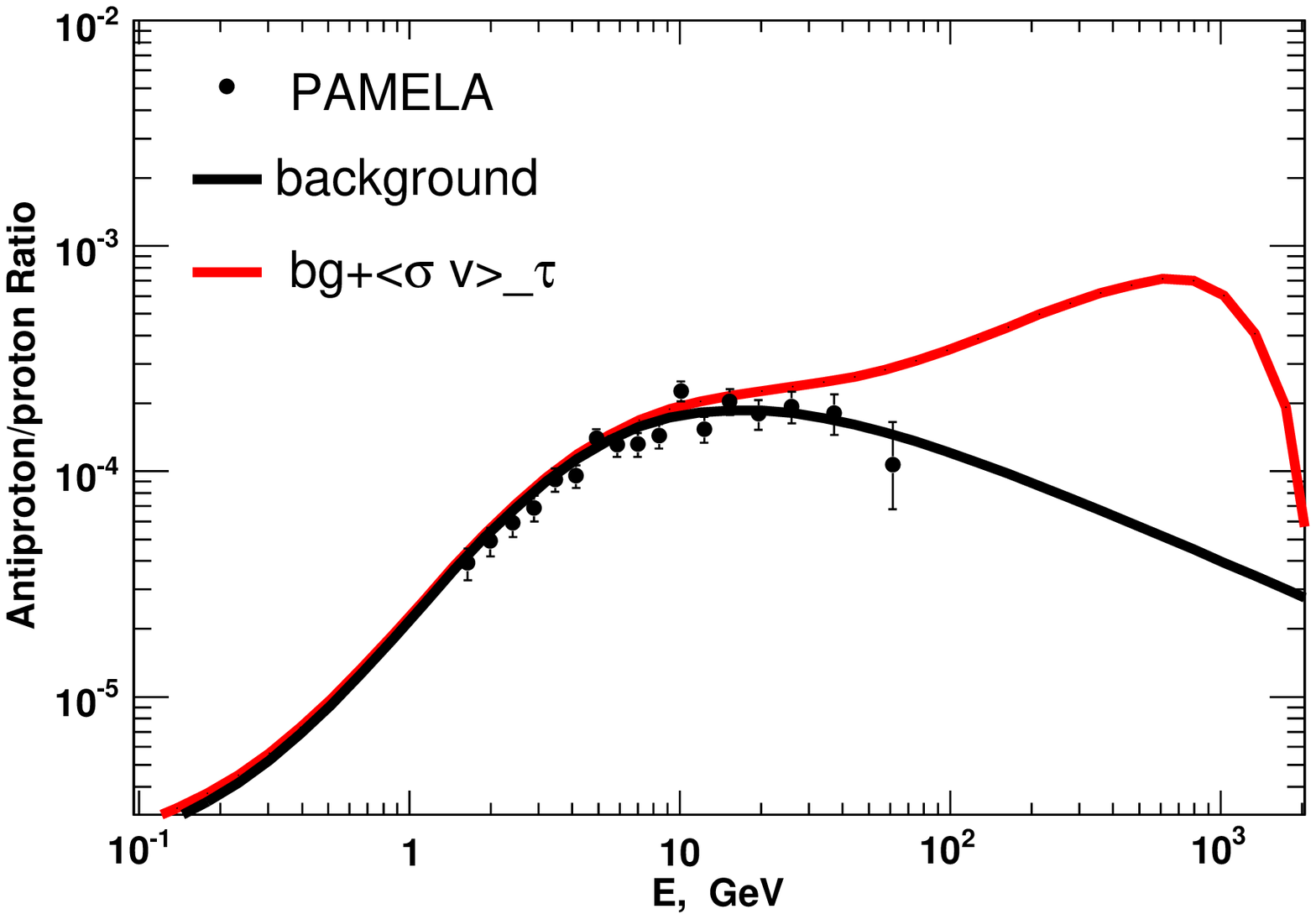}
\caption{The same as Fig. \ref{anti-proton4kpc}
 but for $Z_h=2$kpc.} \label{anti-proton2kpc}
\end{figure}

The quark and anti-quark produced from DM annihilation are subsequently fragmentate into hadrons
and may cause excesses in cosmic proton and anti-proton spectra.
In our analysis the anti-proton spectrum from
quark pair hadornization is treated with the
PYTHIA package\cite{Sjostrand:2006za}.
We find that different quark flavors have
approximately the same  proton and anti-proton spectra in the
final products.  Our results are shown in Fig. \ref{anti-proton4kpc}
and Fig. \ref{anti-proton2kpc} with the diffuse height $Z_h=4$kpc and $2$kpc
respectively.

From the figures we see that at the high energy
end of the anti-proton
spectrum, there is already tension between the theoretical prediction and
the current PAMELA data. For the diffusion hight $Z_h=4$kpc in
Fig. \ref{anti-proton4kpc} we notice that the antiproton fraction is
obviously too much. However, due to large uncertainties in the
propagation models, there are rooms to adjust parameters to fit the
data. When we adopt the diffusion height $Z_h=2$kpc, as shown in Fig.
\ref{anti-proton2kpc},  the anti-proton flux is suppressed and the ratio
is consistent with the PAMELA data at about $2\sigma$ level.
It should be noted that for both $Z_h=4$kpc and $Z_h=2$kpc we have
adjusted the other propagation parameters, such as the diffusion
coefficient $D(E)$, so that all the cosmic ray data are well reproduced
in both models. The DM annihilation cross section are normalized so that
the ATIC or Fermi electron spectrum is well fitted, as shown in
Figs. \ref{e4kpc} and \ref{e2kpc}.

We also notice from Figs. \ref{anti-proton4kpc} and \ref{anti-proton2kpc}
that the contribution of anti-protons from DM annihilation is most important
at several hundred GeV. Therefore future data at higher energies
will give much more stringent constraints on the DM contribution to
anti-protons.
We expect that the near future PAMELA and AMS02 data will play a
crucial role in testing different DM annihilation models.

\section{The $\gamma$-ray radiation from the Galactic Center}

Since we require a very
large DM annihilation cross section to account for the
observational $e^\pm$ CR excesses, this will lead to a strong
$\gamma$-ray radiation from the final lepton states, i.e. via
$\psi \bar \psi \to Z^\prime \to l_i \bar l_i \gamma$.
The cross section for high energy $\gamma$ of energy $E_\gamma$,
with $m_\psi >> m_{l_i}$, is given by \cite{fsr}
\begin{eqnarray}
{d\sigma_{l_i\bar l_i \gamma}\over dE_\gamma} = \sigma_{l_i\bar l_i}{\alpha\over \pi} {1\over E_\gamma}\left [ \ln{s'\over m^2_\mu}-1\right ] \left [ 1 +{s^{'2}\over s^2}\right ]\;,
\end{eqnarray}
where $s = 4 m^2_\psi$ and $s' = 4 m_\psi(m_\psi - E_\gamma)$.
For the $\tau$ final states more $\gamma$-rays will be produced from
the $\tau$ hadronic decays, which will be calculated using PYTHIA
package\cite{fsr}.
We will calculate the $\gamma$-ray emission from the
Galactic Center (GC) in the models under consideration and check if they are consistent with the
HESS observation at the GC \cite{hess}.

As can be seen from Eq. (\ref{gammanu}) that the $\gamma$-ray produced by DM annihilation depends on the square of the DM profile $\rho$ and therefore the result will be sensitive
to the form of $\rho$. In the following calculation we will take two popular dark matter
density profiles, the NFW profile \cite{nfw}
and the Einasto profile \cite{einasto}, to predict the $\gamma$-ray flux for comparisons.
The two density profiles take the forms as
\begin{eqnarray}
\label{pro} \rho_{NFW}(r) \,& =& \,
\frac{\rho_s}{\left({r}/{r_s}\right)
\left(1+{r}/{r_s}\right)^{2}} ~,~\, \nonumber\\
\rho_{Eina}(r) \,& =& \, \rho_0\exp\left[-\frac{2}{\alpha}\left(\frac{r^{\alpha}-
R^{\alpha}_{\odot}}{r^{\alpha}_{s}}\right)\right] ~,~\,
\end{eqnarray}
where
we take the scale radius $r_s = 25$kpc, $\alpha =0.23$ for the Einasto
profile and normalize the local density
as $\rho_\odot =0.3 GeV/cm^3$.
Given the density profile, the $\gamma$-ray flux along a specific
direction can be calculated as given in Eqs. (\ref{gammanu}) and
(\ref{average}).

We calculate the diffuse emission at the GC with the galactic latitude $b$ and longitude $l$
in the region $|b|<0.3^\circ$, $|l|< 0.8^\circ$
and compare the result with the
HESS observation \cite{hess}.
To compare with spectra reported by HESS we have followed the
HESS collaboration
to subtract the background from $0.8^\circ<|b|<1.5^\circ$ \cite{hess}.

\begin{figure}[!ht]
\includegraphics[scale=0.5]{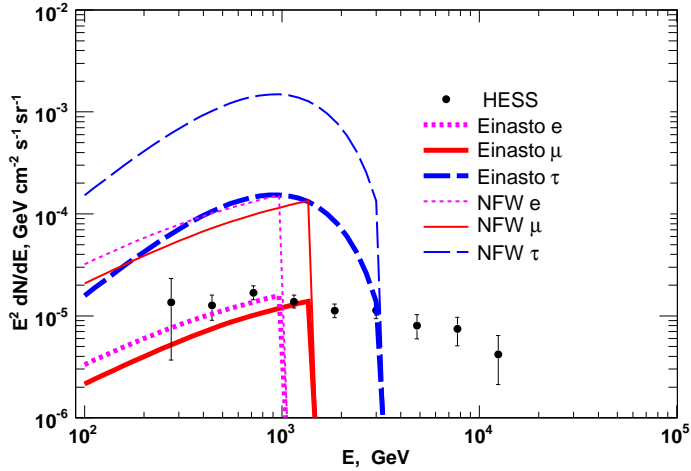}
\caption{The $\gamma$-ray emission from the GC region in the $B-3 L_i$ models.
The data are taken from the HESS observation at the same region \cite{hess}.
We have taken NFW and Einasto profiles to predict the $\gamma$-ray fluxes.
}
\label{gamma-ray}
\end{figure}

In Fig. \ref{gamma-ray} we show the results together with the data
taken from the HESS observation on the same region \cite{hess}.
For the $\tau$ final states we notice the $\gamma$-ray flux
from GC is greatly higher than the observed one. Therefore the
$B-3\tau$ model is disfavored considering its $\gamma$-ray emission
at the GC. Further, $\gamma$-rays also exceed the HESS result
if taking the NFW profile no matter what the final states are taken.
However, the recent high precision simulation show that the density
profile at the halo center does not tend to any simple form of power law
\cite{boylan}.
Instead the Einasto profile gives best fit to the most recent simulations
\cite{boylan,merritt}.
We also notice that if DM annihilates into electrons or muons the
$\gamma$-ray flux is consistent with HESS when taking the Einasto
profile.

It should be noted that for the two propagation models with $Z_h=4$kpc
and $2$kpc the $\gamma$-ray flux from the GC is almost the same, since
the HESS observation focus on a very small region around the GC.

\section{The neutrino cosmic rays }

In the $U(1)_{B-3L_i}$ models, there are large fractions that
DM annihilate into neutrino and
anti-neutrino $\nu_i \bar \nu_i$ pairs directly. The
annihilation rate is given by
\begin{eqnarray}
(\sigma v)_{\nu_i \bar \nu_i} = A {9\over \pi} {a^2 g^{\prime
4}m^2_\psi\over (s - m^2_{Z^\prime})^2 + \Gamma^2_{Z^\prime}
m^2_{Z^\prime}}\; ,\label{neutrino}
\end{eqnarray}
where $A =1$ if the right- and left-handed neutrinos pair up to have Dirac mass, and $A= 1/2$ if the right-handed neutrinos are heavy.
This is a two body annihilation process and the neutrino energy spectrum produced will be line spectrum type.
Therefore the line spectrum neutrinos, if detected, are distinctive signals of
DM annihilation, since no astrophysical processes can produce such
line spectrum.

There are also neutrinos from secondary decays of
$\mu \to \nu_\mu e \bar \nu_e$ for the $U(1)_{B-3L_\mu}$ and
$\tau\to \nu_\tau \mu \bar \nu_\mu + \nu_\tau e \bar \nu_e$ for the
$U(1)_{B-3L_\tau}$ model. $\tau$ can also produce softer neutrinos
by hadronic decays. However, these continuous spectrum neutrinos
are hard to be discriminated from the atmospheric neutrino background.
Further, taking these soft neutrinos into account we have to lower
the threshold of the neutrino detector. The amount of atmospheric
neutrinos increase rapidly by lowering energy since they have a
very soft spectrum \cite{honda}.

There have been some studies of detectability of neutrinos from the GC
by DM annihilation \cite{liujia}. As shown in the last section
the $B-3\tau$ model has been excluded by the $\gamma$-ray emission.
Therefore we calculate the neutrino fluxes in the cases of $B-3e$ and
$B-3\mu$ with the Einasto profile which are consistent with the HESS
$\gamma$-ray observation. We would like to check if the future IceCube
observation can detect the neutrinos predicted by our models at
the GC.

We will consider the muon neutrino detection rate at IceCube.
The flavor eigenstates produced at the source will be modified at the detection point due to neutrino oscillations.
The relation between different flavors
from the GC to the Earth is given by \cite{lin}
\begin{equation}
(\phi_i)_{Earth} = P_{ij} (\phi_j)_{GC}\;,
\end{equation}
where $P_{ij}$ is the probability of converting the j-th flavor type of neutrino at the source to the i-th flavor neutrino at the detection point.

Assuming tri-bimaximal neutrino mixing pattern, the $P_{ij}$ matrix after neutrinos traveling a galactic distance is given by
\begin{eqnarray}
(P_{ij}) = \frac{1}{8}\left(\begin{array}{ccc} 8-4\omega & 2\omega &  2\omega \\
			2\omega & 4-\omega & 4- \omega  \\
  			2\omega & 4-\omega & 4- \omega  \\ \end{array}
\right)\;,
\end{eqnarray}
where $\omega=\sin^2(2\theta_{12})\approx 0.87$. Since the tri-bimaximal mixing is a good approximation to present data, we will use the above formula for
numerical estimate.
We find that  the muon neutrino fluxes
arriving at the Earth are $0.22$ and $0.4$ times of the initial
$\nu_e$ and $\nu_\mu$ fluxes at the sources, respectively.

We consider the detectability of the line neutrino spectrum of our models by the planned
additional DeepCore within the IceCube detector. We take a conservative angular resolution as
large as $\pi$ Sr. Considering the contained events within the DeepCore
it is not difficult to calculate the $\nu_\mu$ events for 4 years
observation with the DeepCore volume 0.04km$^3$. The total
contained muon events are about $606$ for $\nu_{\mu}$ initial flavor
coming from the $\pi$ solid angle centered at
the GC when assuming the DM annihilation cross section is large
enough to account for the cosmic $e^\pm$ anomalies and the Einasto
density profile.
Taking the energy resolution of the detector $\sigma(\log_{10}(E))\sim 0.4$
\cite{resc} (corresponding to the energy bin $600GeV \sim 3.77 TeV$)
and taking the atmospheric neutrino
background of \cite{honda} we get about $2350$ contained background events
at DeepCore/IceCube for 4 years observation from the GC direction
for $\pi$ Sr solid angle.
About $68\%$ signal events will fall within the $1\sigma$ energy range.
Then we find that the DeepCore/IceCube will be able to detect the line spectrum neutrinos
at the confidence level $8.5\sigma$ signal for 4 years observation
with the $B-3\mu$ model, if the DM annihilation rate is fixed by the $e^\pm$ excess data.
For the $B-3e$ model (with the $\nu_e$ initial flavor) the DM mass is
1TeV. Therefore the energy bin is from $400 GeV$ to $2.5 TeV$.
We then get the moun contained events for 4 years observation
are about $243$. The background events are much larger with about $4000$ events
since the energy bin is lower now. In this case we can  only have a $2.6\sigma$ signal for
the $\nu_e$ initial state which will oscillate into $\nu_\mu$.

Here we have taken a conservative angular resolution. Considering
that the DeepCore may have a better angular resolution we may get signals
with much higher significance.
Therefore the neutrino events are very promising signals to test the
DM annihilation model which explain the cosmic $e^\pm$ excesses.
If the scenario is correct we expect to detect
at DeepCore/IceCube of the line spectrum neutrinos predicted in the
$B-3L_i$ models.

\section{Conclusions}

In this work we studied cosmic  $e^\pm$, $\bar p$, $\gamma$ and neutrino rays in the leptocentric $U(1)_{B-3 L_i}$
dark matter models. In these models, DM annihilation into SM particles is mediated by the  $Z^\prime$ gauge boson of
$U(1)_{B-3 L_i}$. The couplings of $Z^\prime$ to leptons are larger than those to quarks leading to larger annihilation of DM
to leptons than hadrons. This naturally explains the $e^\pm$ excesses from PAMELA, ATIC or FERMI observational data, and at the same time these
models can suppress the anti-proton flux efficiently. We find that the $U(1)_{B-3 L_e}$ model can fit PAMELA and ATIC, while
$U(1)_{B-3 \mu}$ and $U(1)_{B-3 \tau}$ can fit PAMELA and FERMI, respectively. Future data can be used to distinguish different models.
We showed that by slightly adjusting
the propagation parameters these models can predict anti-proton flux consistent
with the PAMELA data.
But these models are different than pure leptophilic models which predict negligible
anti-proton cosmic ray in all energy ranges, the leptcentric $U(1)_{B-3L_i}$ models
predict anti-proton flux beyond the background at higher energies which can be tested by the near
future data from PAMELA or AMS02.

The $U(1)_{B-3L_i}$ models have predictions for cosmic $\gamma$-rays from DM annihilation.
We investigated the $\gamma$-ray
emission from the GC in these models. We found that the  $U(1)_{B-3 \tau}$ model
is excluded by the HESS observation of the GC region since this
model predicts too much $\gamma$-rays. Adopting the Einasto DM density profile
the other two models predict $\gamma$-rays consistent with HESS data.

We also investigated the detectability of neutrinos from the GC by
the $U(1)_{B-3 e}$ and $U(1)_{B-3 \mu}$ models at DeepCore/IceCube. In these models the fraction of
the DM annihilation into neutrinos are large and produce line spectrum
neutrinos detectable by DeepCore/IceCube.
We found that the DeepCore/IceCube can have a $8.5(2.6)\sigma$ signal of the line neutrino
spectrum for 4 years data taking. This will be a distinctive signal
of DM annihilation in our leptocentric $U(1)_{B-3e}$ and $U(1)_{B-3\mu}$ models.

Finally we would like to have some comments on the detectability of leptocentric DM model at the LHC and direct DM detection experiments.
Since the leptocentric DM models discussed here are very different than the pure leptophilic DM models, where the $Z^\prime$ only has interactions
with leptons at the tree level, the $Z^\prime$ in our cases has interaction with quarks and may lead to detectable effects at the LHC
and also direct DM search on the Earth. However because we rely on
Breit-Wigner resonant enhancement effect to produce the
large boos factor, the $Z^\prime$  mass is constrained to be 2 times of the DM mass and is constrained to be in the TeV range and the couplings
$a g{^\prime}^2$ to be in the range of $1\times10^{-5}$ to $5\times10^{-5}$,
the cross section for producing $Z^\prime$ and virtual effects due to $Z'$ for the LHC and also for direct detection
are small, it is not possible to have observable effects at the LHC and near future direct DM detection experiments.
To have a lower $Z^\prime$ mass and therefore to have
observable effects at the LHC and direct DM detection experiments, a different mechanism for the boost factor has to be in effective
to relax the relation that the $Z^\prime$ mass is about 2 times of the DM mass.  To this end we note that if there is a long range interaction between
DM by exchanging a light new particle, it is possible to have the Sommerfeld effect in operation, e.g. in Ref. \cite{review,hisano}. This offers another interesting possibility to explain the
cosmic data. We will give details for this possibility elsewhere.

\acknowledgements
We thank Yuan Qiang for helpful discussions in our calculations of
$\gamma$-ray and neutrino emission.
This work was supported in part by the NSF of
China under grant No. 10773011, by the Chinese Academy of
Sciences under the grant No. KJCX3-SYW-N2, by NSC and NCTS, and by the DOE under Grant No. DE-FG03-94ER40837..


\end{document}